\documentclass[review,12pt,numbers]{elsarticle}
\usepackage{tikz}
\usepackage{xcolor}
\usepackage{lipsum}
\usepackage{amsfonts}
\usepackage{graphicx}
\usepackage{subfig}
\usepackage{epstopdf}
\usepackage{algorithmic}

\usepackage[final]{changes} 
\definechangesauthor[color=blue]{R1}
\definechangesauthor[color=violet]{R2}
\definechangesauthor[color=red]{0}

\usetikzlibrary{positioning, shapes.geometric, arrows, calc}

\setcitestyle{square}

\definecolor{set3color1}{RGB}{141, 211, 199}

\usepackage{amssymb}
\usepackage{amsmath}

\journal{Computer-Aided Design}

\begin{document}

\begin{frontmatter}



\title{Optimization of a Triangular Delaunay Mesh Generator using Reinforcement Learning} 


\author[LBLANAG,UCBAST]{Will Thacher\corref{cor1}} 
\ead{wgt3@berkeley.edu}
\cortext[cor1]{Corresponding Author}
\author[LBLMath,UCBMath]{Yulong Pan}
\ead{yllpan@berkeley.edu}
\author[LBLMath,UCBMath]{Per-Olof Persson}
\ead{persson@berkeley.edu}

\affiliation[LBLANAG]{organization={Applied Numerical Algorithms Group, Lawrence Berkeley National Laboratory},
            city={Berkeley},
            state={CA},
            postcode={94720}, 
            country={USA}}

\affiliation[UCBAST]{organization={Applied Science and Technology Group, University of California Berkeley},
            city={Berkeley},
            state={CA},
            postcode={94720}, 
            country={USA}}

\affiliation[LBLMath]{organization={Mathematics Group, Lawrence Berkeley National Laboratory},
            city={Berkeley},
            state={CA},
            postcode={94720}, 
            country={USA}}

\affiliation[UCBMath]{organization={Department of Mathematics, University of California Berkeley},
            city={Berkeley},
            state={CA},
            postcode={94720}, 
            country={USA}}

\begin{abstract}
In this work we introduce a triangular Delaunay mesh generator that can be trained using reinforcement learning to maximize a given mesh quality metric. Our mesh generator consists of a graph neural network that distributes and modifies vertices, and a standard Delaunay algorithm to triangulate the vertices. We explore various design choices and evaluate our mesh generator on \deleted[id=0]{various} \added[id=0]{diverse} tasks including mesh generation, mesh improvement, and producing variable resolution meshes. The learned mesh generator outputs meshes that are comparable \added[id=0]{in quality} to those produced by Triangle and DistMesh, two popular Delaunay-based mesh generators.
\end{abstract}



\begin{keyword}


Mesh Generation \sep Delaunay Triangulation \sep Reinforcement Learning

\end{keyword}

\end{frontmatter}



\section{Introduction}
Unstructured meshes are a critical component of applications such as numerical simulation and computer graphics. The ``quality" of the mesh can have a significant effect on the output, e.g. the accuracy of a numerical simulation. In addition, the mesh generation itself may incur a significant computational expense. The exact definition of the quality of a mesh will vary between applications, and may include metrics like edge lengths, angles, vertex connectivity, element volumes and element shapes. While there are many mesh generation algorithms that produce high quality meshes by some definition, most of these algorithms do not attempt to maximize a particular mesh metric explicitly \cite{peraire1987adaptive, owen1999q,remacle2012blossom,shewchuk2002delaunay}. 

One reason for this is the difficulty of posing and solving an appropriate optimization problem. Suppose we have selected a quality metric which assigns any mesh a score as a function of vertex positions and connectivities. Maximizing this mixed-integer optimization problem over the space of meshes is far from straightforward, as it involves both discrete and continuous components: one must search over the number and position of vertices, as well as which vertices are connected by edges. Furthermore, it may be unreasonable to solve a full optimization problem each time we are given a new domain to mesh. 

Several authors \cite{NARAYANAN2024103744, PAN2023288,lei2023whatssituationintelligentmesh} have addressed these difficulties by viewing mesh generation as a sequence of decisions that update a mesh from an initial to a final state. Rather than searching for optimal vertex positions and connectivities directly, we can instead try to find an optimal decision making strategy that produces a high quality mesh, in the chosen metric, for any given domain. If we can define a parameterized decision making strategy, the optimization problem then becomes to maximize the quality of meshes produced by this decision making process over the space of parameters. In other words, we will ``train" the mesh generator itself. It is not immediately apparent that this will simplify the optimization problem, because the decision making strategy will still involve both continuous and discrete actions. However, this approach opens us up to utilizing reinforcement learning techniques that have been specifically designed for such optimal decision making problems.

In this paper we propose a mesh generator that can be trained using reinforcement learning, and apply it to the problem of generating triangular meshes on polygonal domains in two dimensions. In brief, our mesh generator consists of two components: a parameterized function that moves, adds and deletes vertices, and the Delaunay algorithm \cite{Ruppert1995ADR}, which produces a triangulation of these vertices. The mesh generator alternates between vertex modification and triangulation for some number of iterations. The vertex modification function is a graph neural network, which takes advantage of the natural graph structure of meshes.

In Section \ref{sec:MG} we describe our mesh generator and the reinforcement learning algorithm we use to optimize it. In Section \ref{sec:eval} we experiment with different design choices and evaluate our mesh generator, comparing it to popular Delaunay-based mesh generators.

\subsection{\added[id=R1]{Prior Work} }
\added[id=R1]{Prior works in this area can be differentiated based on 1) which aspect of mesh generation they are trying to improve, 2) the formulation of a decision making process to do so, 3) the optimization objective, and 4) the training algorithm. In \cite{NARAYANAN2024103744}, the authors address the problem of improving the quality of existing triangular or quadrilateral meshes. They formulate a decision making process which involves purely edge-based actions such as flipping and splitting, and train their mesh editor to optimize the degree of each vertex, or the number of vertices incident on each node. In \cite{PAN2023288}, a fully automated end-to-end quadrilateral mesh generator is developed. The decision making process includes placing new vertices to create quadrilateral elements one at a time, moving inward from the boundary. They optimize for element quality and volume. The works of \cite{Lorsung} and \cite{adaptive} focus on the problem of adaptively modifying meshes during complex physics simulations. In \cite{Lorsung}, a vertex removal strategy is learned, and in \cite{adaptive} an element refinement strategy is learned. Both utilize graph neural networks to parameterize the decision making strategy, and the optimization objective in both is to minimize simulation error. These four papers use a variety of state-of-the-art training algorithms including Proximal Policy Optimization \cite{schulman2017proximalpolicyoptimizationalgorithms}, soft Actor-Critic \cite{haarnoja2018softactorcriticoffpolicymaximum}, and Deep Q Learning \cite{van2016deep}. There are other works which use various machine learning techniques, but not necessarily reinforcement learning, to improve some part of mesh generation; a good overview is available in the introduction of \cite{PAPAGIANNOPOULOS2021152} }.

\added[id=R1]{Our approach differs from prior works in several key ways. Firstly, our decision making process is flexible, as any existing vertex can be moved or deleted, and new vertices can be added. This allows our trained mesh generator to both create new meshes and improve existing meshes. Secondly, to our knowledge it is the first work in this area that uses a graph neural network for the purpose of building high quality meshes. By treating the mesh as a graph we can formulate a global decision making process which involves every vertex of the mesh, which is different from the local action spaces used in \cite{NARAYANAN2024103744} and \cite{PAN2023288}.
}

\section{Mesh Generator} \label{sec:MG}
Our objective is to train a mesh generator to produce high quality triangular meshes on two dimensional polygonal domains. The boundary of the domain $\Omega \subset \mathbb{R}^2$ is described as an ordered list of vertices. The output of the mesh generator is a list of vertices $\{x_i\}$ and a list of triangles $\{T_k\}$, where each triangle consists of a tuple of three vertices $[x_{k_1},x_{k_2},x_{k_3}]$. In this section we will describe our mesh generator using reinforcement learning terminology.

\subsection{Algorithm} \label{sec:MG_ALG}
Let us denote the initial set of polygon vertices as $\{x_i^0\}$. Given $\{x_i^0\}$, we can call the Delaunay algorithm to produce an initial triangulation $\{T_k^0\}$. This triangulation is naturally represented as a graph $ (\mathcal{V}^0,\mathcal{E}^0)$, where $\mathcal{V}^0:=\{x_i^0\}$ are the spatial coordinates of the vertices, or nodes, and $e_{ij} \in \mathcal{E}^0$ means there is a mesh edge between nodes $i$ and $j$. The graph $s^t := (\mathcal{V}^t,\mathcal{E}^t)$ is the \emph{state} of the mesh at timestep $t$.

The decision making strategy is realized numerically as a \emph{policy}  function $\pi_{\theta}$ which maps a state $s^t$ to an \emph{action} $a^t$.  The \emph{action space} is as follows:
\begin{enumerate}
    \item Each non-boundary node may be moved
    \item Each non-boundary node may be deleted
    \item A new point may be added at the midpoint of any existing edge, including boundary edges
    \item A new point may be added at the centroid of any existing triangle
\end{enumerate}
\added[id=R2]{We note here that the action space is dependent on the current state, unlike some reinforcement learning problems in which the action space is fixed.} For optimization purposes, the policy will be stochastic, meaning it maps $s^t$ to a probability distribution over actions that can be sampled from. This combination of moving, adding, and deleting points each iteration is also used in the two dimensional meshing algorithm described in \cite{gmesh}. Further details about the policy function are given in a subsequent section. 

The \emph{environment} decodes the numerical output of the policy function into actions that can update the current state, assesses the validity of the proposed actions, and updates the vertex list to obtain $\{x_i^1 \}$. The environment then calls the Delaunay algorithm to produce a triangulation of the current node positions, giving us a new state $s^1$. The sequence of states $s^0,s^1...s^n$ comprise a \emph{trajectory} $\tau$. Because the policy is stochastic, given an initial state $s^0$ the trajectories will be distributed according to some probability distribution $p_{\theta}(\tau)$.

\begin{figure}

\subfloat[]{
\begin{tikzpicture}
  \coordinate (A) at (0, 0);
  \coordinate (B) at (4, .5);
  \coordinate (C) at (3, 3);
  \coordinate (D) at (1.5, 4);
  \coordinate (E) at (-1, 3);
  \coordinate (F) at (-2, 1);
  
  \coordinate (G) at (1, 1.5);
  \coordinate (H) at (0, 2.5);
  \coordinate (I) at (2.5, 1.5);
  \coordinate (J) at (2, 2);

  \fill[set3color1] (A) -- (B) -- (C) -- (D) -- (E) -- (F) -- cycle;
  \draw[thick] (A) -- (B) -- (C) -- (D) -- (E) -- (F) -- cycle;

  \draw[thick] (A) -- (G);
  \draw[thick] (B) -- (G);
  \draw[thick] (B) -- (I);
  \draw[thick] (C) -- (I);
  \draw[thick] (C) -- (H);
  \draw[thick] (D) -- (H);
  \draw[thick] (E) -- (H);
  \draw[thick] (F) -- (G);
  \draw[thick] (G) -- (H);
  \draw[thick] (G) -- (J);
  \draw[thick] (I) -- (G);
  \draw[thick] (H) -- (F);

  \draw[thick] (J) -- (H);
  \draw[thick] (J) -- (C);
  \draw[thick] (J) -- (I);

  \foreach \point in {A, B, C, D, E, F} {
    \fill[red] (\point) circle (2pt);
  }

  \foreach \point in {G, H, I,J} {
    \fill[black] (\point) circle (2pt);
  }

  \node[below, yshift=-4pt] at (G) {$x_1$};
  \node[above, yshift=2pt] at (H) {$x_2$};
  \node[right] at (I) {$x_3$};
  \node[above, yshift=2pt] at (J) {$x_4$};
\end{tikzpicture}
}
\subfloat[]{
\begin{tikzpicture}
  \coordinate (A) at (0, 0);
  \coordinate (B) at (4, .5);
  \coordinate (C) at (3, 3);
  \coordinate (D) at (1.5, 4);
  \coordinate (E) at (-1, 3);
  \coordinate (F) at (-2, 1);

  \coordinate (G) at (.25, 1.6);
  \coordinate (H) at (1.5, 2.5);
  \coordinate (I) at (2.1, 1.2);
  \coordinate (J) at (2, 2);

  \fill[set3color1] (A) -- (B) -- (C) -- (D) -- (E) -- (F) -- cycle;
  \draw[thick] (A) -- (B) -- (C) -- (D) -- (E) -- (F) -- cycle;


  \foreach \point in {A, B, C, D, E, F} {
    \fill[red] (\point) circle (2pt);
  }

  \foreach \point in {G, H, I} {
    \fill[black] (\point) circle (2pt);
  }

  \foreach \point in {J} {
    \fill[blue] (\point) circle (2pt);
  }

  \node[below] at (G) {$x_1$};
  \node[above] at (H) {$x_2$};
  \node[right] at (I) {$x_3$};
  \node[above] at (J) {$x_4$};

  \node[below left] at (A) {
    $\color{red}\alpha_e = .1$
  };

  \node[below left] at (B) {
    $\color{red}\alpha_e = .2$
  };

  \node[below right] at (J) {
    $\color{blue}\alpha_d = .3$
  };
\end{tikzpicture}
}
\hspace{0mm}
\subfloat[]{
\begin{tikzpicture}
  \coordinate (A) at (0, 0);
  \coordinate (B) at (4, .5);
  \coordinate (C) at (3, 3);
  \coordinate (D) at (1.5, 4);
  \coordinate (E) at (-1, 3);
  \coordinate (F) at (-2, 1);
  
  \coordinate (G) at (.25, 1.6);
  \coordinate (H) at (1.5, 2.5);
  \coordinate (I) at (2.1, 1.2);
\coordinate (J) at (2, .25);

  \fill[set3color1] (A) -- (B) -- (C) -- (D) -- (E) -- (F) -- cycle;
  \draw[thick] (A) -- (B) -- (C) -- (D) -- (E) -- (F) -- cycle;


  \foreach \point in {A, B, C, D, E, F} {
    \fill[red] (\point) circle (2pt);
  }

  \foreach \point in {G, H, I} {
    \fill[black] (\point) circle (2pt);
  }

  \foreach \point in {J} {
    \fill[orange] (\point) circle (2pt);
  }


\end{tikzpicture}
}
\subfloat[]{
\begin{tikzpicture}
  \coordinate (A) at (0, 0);
  \coordinate (B) at (4, .5);
  \coordinate (C) at (3, 3);
  \coordinate (D) at (1.5, 4);
  \coordinate (E) at (-1, 3);
  \coordinate (F) at (-2, 1);
  
  \coordinate (G) at (.25, 1.6);
  \coordinate (H) at (1.5, 2.5);
  \coordinate (I) at (2.1, 1.2);
\coordinate (J) at (2, .25);

  \fill[set3color1] (A) -- (B) -- (C) -- (D) -- (E) -- (F) -- cycle;
  \draw[thick] (A) -- (B) -- (C) -- (D) -- (E) -- (F) -- cycle;

  \draw[thick] (A) -- (G);
  \draw[thick] (G) -- (J);
  \draw[thick] (B) -- (I);
  \draw[thick] (C) -- (I);
  \draw[thick] (C) -- (H);
  \draw[thick] (D) -- (H);
  \draw[thick] (E) -- (H);
  \draw[thick] (F) -- (G);
  \draw[thick] (G) -- (H);
  \draw[thick] (I) -- (H);
  \draw[thick] (I) -- (G);
  \draw[thick] (E) -- (G);
  \draw[thick] (I) -- (J);

  \foreach \point in {A, B, C, D, E, F} {
    \fill[red] (\point) circle (2pt);
  }

  \foreach \point in {G, H, I,J} {
    \fill[black] (\point) circle (2pt);
  }


\end{tikzpicture}
}
\caption{(a) Given an initial triangulation, the mesh state $s^t$ is passed through the policy function. (b) Updated node positions and action variables are sampled from $\pi_{\theta}(a^t|s^t)$. (c) Nodes are moved, added and deleted accordingly. The node in blue is deleted and the orange node is added at the midpoint of an existing edge. (d) The vertices are retriangulated to produce a new mesh. }
\label{fig:mesh-gen}
\end{figure}
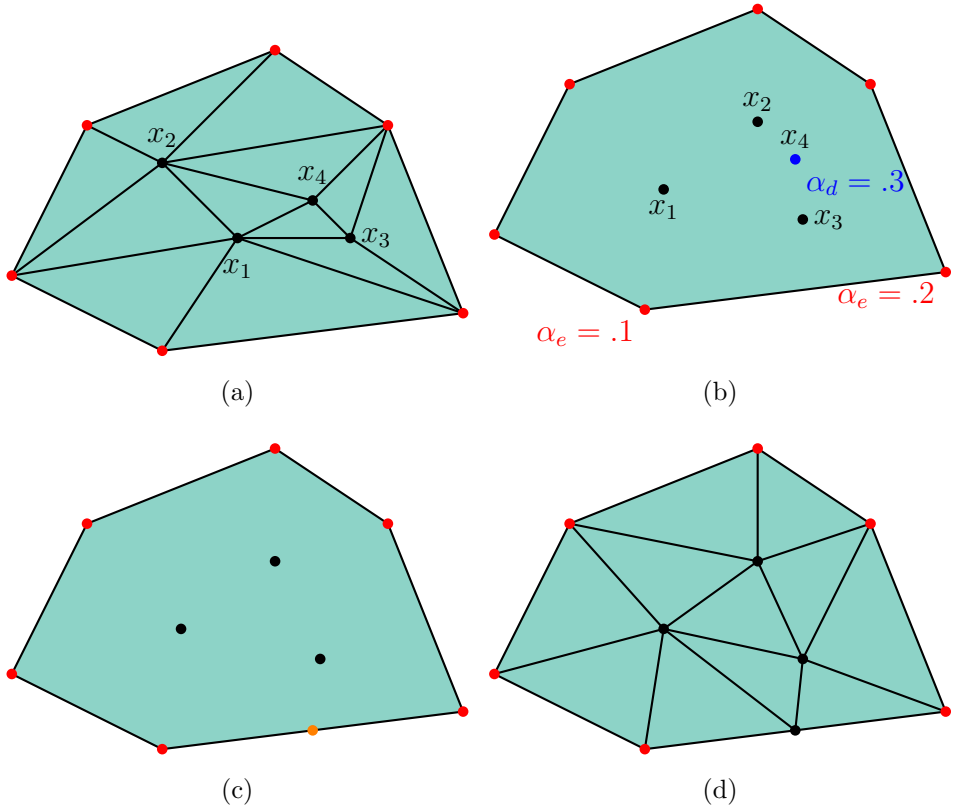

To determine if a particular action or series of actions is desirable, we define a \emph{reward} function $r^t$. Given a mesh state $s^t$, we record four ``quality" metrics which measure the deviation of various geometric quantities from target values. Meshes are normalized so that the target edge length is 1 and the target element volume is $\sqrt{3}/4$, the area of an equilateral triangle with side length 1.
\begin{enumerate}
    \item Edge length quality: $q_e (e_{ij}) = 1 - |1 - |e_{ij}||$, where $|e_{ij}| = \Vert x_i - x_j \Vert $
    \item Angle quality: $q_a (\gamma_l ) = 1 - \frac{|\gamma_l^* - \gamma_l|}{\gamma_l^*}$, where $\gamma_l$ is the measurement of angle $l$ and $\gamma_l^*$ is the desired angle for that vertex
    \item Volume quality: $q_v (T_k) = 1 - \frac{\left| |T_k| - \sqrt{3}/4 \right|}{\sqrt{3}/4}$, where $|T_k|$ is the volume of element $k$
    \item Element quality:  $q_r (T_k)  = 2 \frac{\rho_{k,\text{out}}}{\rho_{k,\text{in}}}$, where $\rho_\text{out}$ is the radius of the smallest circumscribed circle and $\rho_\text{in}$ is the radius of the largest inscribed circle of element $T_k$ \cite{field00quality}
\end{enumerate}
Let $S(s^t)$ be a total score for the mesh, which is a weighted average of some discrete norm of each quality metric over the mesh. The reward for a single timestep is defined as $r^t(s^t,a^t,s^{t+1}):=S(s^{t+1})-S(s^t)$, and the reward $R$ for a trajectory $\tau$ is defined as $R(\tau) := \sum_{t=1}^n r^t$ and we set $S(s^0) = 0$ and do not use any discount factor so that $R(\tau) = S(s^n)$, the final score of the mesh. 

We will train our mesh generator using the following objective:
\begin{align}
    \max_{\theta} \ \mathbb{E}_{\Omega \sim p_D} \left[ \mathbb{E}_{\tau \sim p_{\theta}} R(\tau) \right]
\end{align}
where the polygonal domains are sampled from some distribution $p_{D}$\added[id=0]{, and $\theta$ are the trainable parameters of the policy function}. The inner expectation can be maximized using a \emph{policy gradient} method. \cite{hardtrecht2022patterns}

\subsection{Policy Function}
\begin{figure}
    \centering
    \includegraphics[width=1\linewidth]{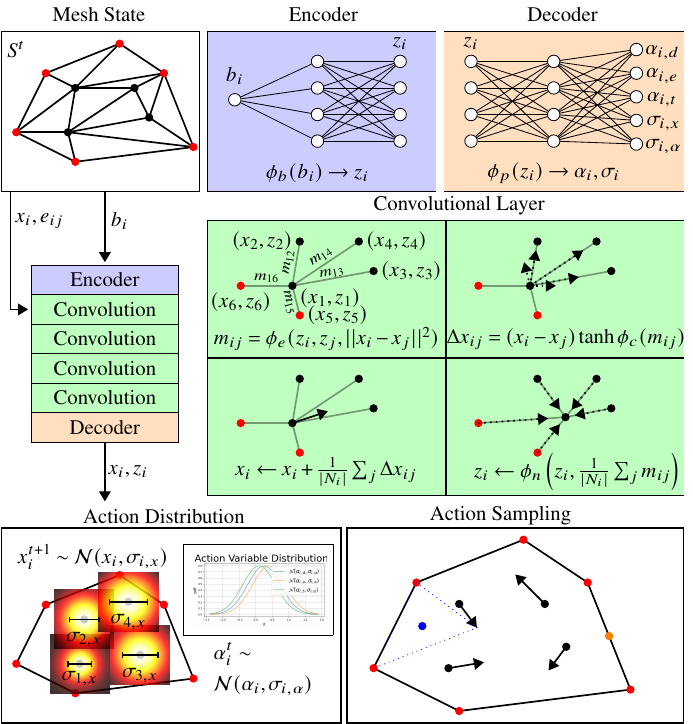}
    \caption{\added[id=R2]{Policy network architecture. Binary labels are mapped to feature vectors by a multi-layer perceptron (MLP) in the \textbf{encoder} layer. The \textbf{convolutional} layer is visualized in four parts from top left to bottom right: 1) Feature vectors and inter-node distances are mapped to vectors $m_{ij}$ on each edge as in \eqref{eq:dist_update}. 2) Node position updates are calculated from \eqref{eq:x_update_form}. 3) Nodes are repositioned according to \eqref{eq:x_update}. 4) Feature vectors are updated using \eqref{z_update}. The \textbf{decoder} layer maps features to distribution parameters. The \textbf{action distribution} is comprised of 2D Gaussian distributions for new interior node positions and 1D Gaussian distributions for action variables. New node positions and new node placements are obtained by sampling from this distribution, with the black arrows indicating interior node movement. The blue node is added at the centroid of a triangle and the orange node is added at the centroid of an edge based on the numerical value of the sampled action variables at nodes.} }
    \label{fig:GNN}
\end{figure}

The policy function maps a state $s_t$ to a probability distribution $\pi_{\theta}(a_t | s_t)$ over possible actions. \deleted[id=0]{Our policy function should meet several criteria. Firstly, the input to the policy can be a graph of arbitrary size, so ideally the computational cost should scale linearly with the number of vertices in the mesh. Secondly, it should be straightforward to compute $\nabla_{\theta} \pi_{\theta}(a_t|s_t)$.} \added[id=R2]{The input to this function will be a graph of arbitrary size and connectivity, and the outputted probability distribution will be defined by distribution parameters at each node of the graph. For optimization purposes we must be able to straightforwardly compute $\nabla_{\theta} \pi_{\theta}(a_t|s_t)$. } To meet these \deleted[id=0]{two} requirements, the policy function will be a graph neural network (GNN) with convolutional layers, for which required gradients can be computed using automatic differentiation. \added[id=R2]{For readers unfamiliar with convolutional graph neural networks, an accessible introduction is available from \cite{sanchez}. Our GNN consists of an ``encoder" layer, a series of convolutional layers, and a ``decoder" layer. Over these layers the GNN updates a graph data structure which has the same connectivity as the input state, and stores a vector of values at each node which we refer to as ``feature" vectors. These feature vectors are decoded into the outputted distribution parameters. The policy function pipeline is visualized in Figure \ref{fig:GNN}.}

\added[id=R2]{The purpose of the encoder layer is to initialize feature vectors at each vertex that will be updated throughout the convolutional layers. The vertices are given binary labels $b_i$ indicating whether or not they lie on the boundary of $\Omega$. The encoder layer consists of a multi-layer perceptron (MLP) $\phi_b$ which maps each label $b_i$ to a higher dimensional feature vector $z_i$. Thus the initial feature vector at boundary nodes will be $\phi_b(0)$ and the initial feature vector at non-boundary nodes will be $\phi_b(1)$. }

\deleted[id=0]{The input feature $b_i$ is a binary label which is 1 if $b_i$ lies on the boundary of $\Omega$. That feature is fed through a pointwise ``encoder" $\phi_b$, which is a multi-layer perceptron (MLP) whose output is a higher dimensional vector $z_i$.} \deleted[id=0]{Our GNN} \added[id=0]{The convolutional layer} architecture is an adaptation of a symmetry preserving GNN introduced in \cite{satorras2022enequivariantgraphneural}. \added[id=R2]{In this layer, node positions and feature vectors are updated using information at neighboring nodes. The neighborhood $N_i$ of node $i$ consists of all nodes $j$ for which the edge $e_{ij}$ exists. Let $\phi_e, \phi_c, \phi_n$ denote MLPs, where $e,c,n$ stand for ``edge," ``coordinate" and ``node" respectively. The update equations in the convolutional layer are as follows:} \deleted[id=0]{The node positions and feature vectors are updated through several convolutional layers of the following form:}

\begin{align}
    m_{ij} &= \phi_e(z_i,z_j,||x_i-x_j||^2) \label{eq:dist_update} \\
    \Delta x_{ij} &=  (x_i - x_j) \tanh{\phi_c(m_{ij})} \label{eq:x_update_form} \\
    x_i &\gets x_i + \mathbf{1}_0(b_i) \frac{1}{|N_i|} \sum_{j \in N_i}\Delta x_i\label{eq:x_update}  \\
    z_i &\gets \phi_n \left(z_i, \frac{1}{|N_i|} \sum_{j \in N_i} m_{ij}\right) \label{z_update}
\end{align}

\deleted[id=0]{where $\phi_e, \phi_c, \phi_n$ are MLPs ($e,c,n$ stand for ``edge," ``coordinate" and ``node" respectively).} \added[id=R2]{$\phi_e$ maps features $z_i$ and $z_j$ and the internode distance to a vector $m_{ij}$ which has the same dimension as $z_i$. $\phi_c$ maps $m_{ij}$ to a scalar which is passed through a $\tanh$ function so that this output lies in $(-1,1)$. We multiply this output by the vector $(x_i-x_j)$ and average over $N_i$ to obtain an update vector $\Delta x_i$ which is added to the position of each non-boundary node. Finally, we update the feature $z_i$ using the MLP $\phi_n$, which maps the current feature $z_i$ and the average of the $m_{ij}$ over $N_i$. } \deleted[id=0]{This convolutional layer is visualized in Figure \ref{fig:GNN} The indicator function $\mathbf{1}_0(b_i)$ is used to prevent the boundary nodes from moving. $m_{ij}$ transmits information from node $j$ to node $i$, and in the GNN literature it is known as a ``message." It has the same dimension as $z_i$.}This architecture bears some similarity to the update function in DistMesh, \cite{distmesh} in which nodes are repositioned as a function of inter-neighbor distances. It can be checked that if we rotate and translate the input coordinate positions, the output coordinates $x_i$ will be rotated and translated by the same amount, and the outputted features $z_i$ will not change. Another feature of this architecture is that if an interior vertex $i$ is a member of six interior equilateral triangles, we will have $\Delta x_i = 0$ because the incoming $m_{ij}$ will be identical and the updates from neighbors opposite each other will cancel. Thus a perfect mesh will not be disturbed over a single convolutional layer.

\deleted[id=R2]{Lastly, $z_i$ is passed through a ``decoder" MLP $\phi_p$ and a sigmoid function} \added[id=R2]{The decoder layer consists of an MLP $\phi_p$ which maps each feature $z_i$ to a five dimensional vector which is then passed through a sigmoid function so that each component of the vector is the range $(0,1)$. The five components of this vector, as well as the final node positions, will be the parameters of our probability distribution over actions. The first three components represent} \deleted[id=R2]{to obtain} ``action" variables which we will denote as $\alpha_{i,d}, \alpha_{i,e}, \alpha_{i,t} $, \deleted[id=R2]{as well as two} \added[id=R2]{and the last two are} variance variables $\sigma_{i,x}, \sigma_{i,\alpha}$. The final policy distribution is the product of independent normal distributions at each node with mean $[x_i, \alpha_{i,d}, \alpha_{i,e}, \alpha_{i,t} ]$ and variance $[\sigma_{i,x}^2 \mathbf{I}_2, \sigma_{i,\alpha}^2  \mathbf{I}_3]$. 

\deleted[id=R2]{Using sampled node positions and action variables, the environment will update the mesh as follows:} \added[id=R2]{We sample from this distribution to obtain new node positions and action variables. With a slight abuse of notation we also here denote the sampled action variables and node positions as  $x_i, \alpha_{i,d}, \alpha_{i,e}, \alpha_{i,t}$, which will be used to update the mesh as follows:}
\begin{enumerate}
    \item If $x_i$ is located within $\Omega$, set \deleted[id=0]{$x_i^1 = x_i$; otherwise keep $x_i^1 = x_i^0$} \added[id=0]{$x_i^{t+1} = x_i$; otherwise keep $x_i^{t+1} = x_i^t$}
    \item If $\alpha_{i,d} \leq \nu$, delete node $i$
    \item For all existing edges $e_{ij}$, if $\frac{1}{2}(\alpha_{i,e}+\alpha_{j,e}) \leq \nu$, and $\frac{1}{2}(x_i^1 + x_j^1) \in \Omega$, add that point as a new vertex
    \item For all existing triangles $t_k^0$, if $\frac{1}{3}(\alpha_{k_1,t}+\alpha_{k_2,t}+\alpha_{k_3,t}) \leq \nu$, and if $\frac{1}{3}(x_{k_1}^1 + x_{k_2}^1+x_{k_3}^1) \in \Omega$, add that point as a new vertex
\end{enumerate}
where $\nu$ is some chosen parameter; we use $\nu=\frac{1}{4}$. All MLPs in the network have one hidden layer \added[id=0]{and use a Swish activation function}. See Figure \ref{fig:mesh-gen} for a visual example of one step of a trajectory.

The stochastic policy allows us to explore the action space and optimize the parameters of the probability distribution $\pi_{\theta}(a^t|s^t)$ so that trajectories with high rewards become likely. This is the general idea behind policy gradient methods; see \cite{hardtrecht2022patterns} for an introduction. Details of the optimization algorithm are given in the next section. After training is completed, when running the policy function we have the option to use the average of each distribution as the output, in which case the policy will be deterministic.


\subsection{Optimization Algorithm}
We use the Proximal Policy Optimization (PPO) algorithm \cite{schulman2017proximalpolicyoptimizationalgorithms}, which is an iterative process that alternates between sampling trajectories and updating the policy. We review a few details here in order to cover particulars of our implementation. 

At the start of each PPO iteration we sample a batch of trajectories by running the mesh generator using the current policy function on randomly initialized polygonal domains. \added[id=0]{The specifics of this initialization procedure are given in Section \ref{sec:eval}.} We then form an objective function that will be partially maximized over $\theta$ using a gradient based optimizer. We need a few definitions to specify this objective function.

The \emph{advantage} function is defined as:
\begin{align}
    A_{\theta}(s^t,a_0^t) &= \mathbb{E}_{ a^t = a_0^t} \left[ \sum_{l=0}^\infty r^{t+l}\right] - \mathbb{E}_{a^t \sim \pi_{\theta}(s^t)} \left[ \sum_{l=0}^\infty r^{t+l} \right] \\
    &=: Q(s^t,a_0^t) - V(s^t)
\end{align}
where $r^{t+l} = r^{t+l}(s^{t+l+1}, a^{t+l}, s^{t+l})$. The first term in the right hand sum is the \emph{state-action value}  function $Q$, and the second term is the \emph{value} function $V$. The advantage function measures the difference in expected reward over the remainder of the trajectory between choosing the action $a_0^t$ versus using the average action sampled from the current policy as a function of $s^t$, $\pi_{\theta}(s^t)$. Rather than maximize the expected reward of an action, we will maximize the expected advantage of an action. We will denote an estimate of the advantage as $\hat{A}_{\theta}$. 

We use the Generalized Advantage Estimation \cite{schulman2018highdimensionalcontinuouscontrolusing} technique, which requires a function that can approximately compute values of arbitrary states. We model the value function as another graph neural network $V_{\phi}$ with parameters $\phi$ which has the same architecture as the policy function, with a few modifications to produce a scalar output. After the convolutional layers, the node features $z_i$ are passed through a node-wise decoder MLP, averaged over the mesh, and passed through one final MLP whose output is a single number. Although it is quite possible to share parameters between the value and policy functions, we have obtained better results by using two separate networks. Each PPO iteration we update $\phi$ by fitting the value function to Monte Carlo estimates of state values provided by the trajectory rewards. For further details we refer to \cite{schulman2018highdimensionalcontinuouscontrolusing}.

Let $\theta_m$ be the parameters at the beginning of iteration $m$ of the PPO algorithm. The preliminary objective function $L$ will be defined as:
\begin{align}
    \max_{\theta} L(\theta) = \mathbb{E}_{\tau \sim p_{\theta_m}} \sum_{t=0}^n {\frac{\pi_{\theta}(a^t|s^t)}{\pi_{\theta_m} (a^t|s^t)} \hat{A}_{\theta_m} (a^t,s^t) } =: \mathbb{E}_{\tau \sim p_{\theta_m}} \sum_{t=0}^n \beta^t(\theta) \hat{A}_{\theta_m} (a^t,s^t) 
\end{align}
Our collected trajectories provide a Monte Carlo estimate of $L$. By maximizing this objective, we adjust the parameters $\theta$ so that actions with positive advantages become more likely (relative to the previous policy) and actions with negative advantages become less likely. The PPO algorithm modifies this objective by ``clipping" the probability ratio to limit the size of the policy update:
\begin{align}
    \max_{\theta} \ L_{PPO}(\theta) = \ \mathbb{E}_{\tau \sim p_{\theta_m}} \sum_{t=0}^n {\min \left(\beta^t(\theta) \hat{A}_{\theta_m}, \hbox{clip}(\beta^t(\theta), 1+\epsilon, 1-\epsilon) \hat{A}_{\theta_m} \right)}
\end{align}
where $\epsilon > 0$ is a chosen parameter. One way to interpret this expression is that we are not incentivized to change the probability of a particular action by any more than a factor of $1 + \epsilon$, with the intention of stabilizing the learning process. In practice we do not fully try to maximize this objective, but instead run some number of epochs of the Adam optimizer \cite{kingma2017adammethodstochasticoptimization} to update $\theta$. Lastly, we add to the objective function an \emph{entropy bonus} term which is meant to encourage ``exploration" of the action space by incentivizing the policy distribution to have higher variance. To be precise, this term is the average over all sampled timesteps (for that PPO iteration) of the entropy of the distribution $\pi_{\theta}(a^t|s^t)$. The entropy of $\pi_{\theta}(a^t|s^t)$ is the sum of the entropies of the normal distributions at each node which are used to sample coordinates and action variables.


The algorithm is implemented using the RLLIB library \cite{liang2018rllibabstractionsdistributedreinforcement}, which provides a parallel implementation of the PPO algorithm with GAE, and we additionally use graph neural network tools from the PyTorch Geometric library \cite{fey2019fastgraphrepresentationlearning}. We use single precision arithmetic; experiments showed that using double precision did not improve performance.

\section{Results} \label{sec:eval}
In this section we experiment with design choices and evaluate our mesh generator. The mesh generator is evaluated on a test set of 50 randomly generated polygonal domains with between 10-40 sides, scaled by a factor of 10, with a trajectory length of 15\added[id=R2]{, a sufficient number of steps for the node positions to converge}. We compare our learned mesh generator with two popular mesh generators that also rely on the Delaunay algorithm: Triangle \cite{shewchuk96b} and DistMesh \cite{distmesh}. When using Triangle we input the boundary vertices, and use \deleted[id=0]{a} \added[id=0]{the following options: the ``quality" flag \textbf{-q}, a} maximum element size of .5\deleted[id=0]{ with}\added[id=0]{, and} an angle quality constraint of $20$ degrees. When using DistMesh, we input the boundary vertices and specify an initial minimum edge length of one. \added[id=R1]{These are standard configurations for these algorithms.} We note here that for these particular domains this may not be the optimal way to use Triangle and DistMesh, as we do not let these mesh generators delete nodes on the boundary which are not vertices of the polygonal domain. However, our mesh generator is not allowed to delete those vertices either, so we consider this to be a fair comparison. We report the mean \emph{over the test set} of the mean, minimum, and standard deviation of each quality metric \emph{over each test mesh}.  The test set is large enough that the standard deviation of each reported statistic is less than ten percent.

The mesh quality measures defined in \ref{sec:MG_ALG} are redundant in various ways; for example, if a triangle has three edges of unit length, it must have perfect angles, radius ratio, and volume. \added[id=R1]{It will be useful for us to define a default score function so that we can examine the effect of changing other variables, such as model size or task, in a controlled manner. In Section \ref{sec:mod} we experiment with different score functions. Our default score function equally weights edge and angle quality:} \deleted[id=R1]{As a starting place, we choose a score function which includes only angles and edge lengths:}
\begin{align}
    S(s^t) = \frac{1}{2|n_e|}\sum_{e_{ij}} q_e(e_{ij}) + \frac{1}{2|n_a|} \sum_{\gamma_l} q_a(\gamma_l) \label{eq:rew}
\end{align}
where $|n_e|$ is the number of non-boundary edges and $|n_a|$ is the number of angles. We do not include boundary edges in the score because they cannot be modified. For an angle whose vertex is not on the boundary, the target angle is $\pi/3$. For an angle whose vertex is on the boundary, a 60 degree angle will not always be achievable, and we define the target angle for that vertex as follows: let $\gamma$ be the angle formed by the polygonal boundary, the target angle will be $\gamma/k$, where the integer $k$ is such that $\gamma/k$ is closest to $\pi/3$. For most polygonal domains, a perfect score will not be achievable.

We employ a learning ``curriculum" in which we increase the size and complexity of the random polygonal domains and the number of trajectory steps, and also modify hyperparameters \deleted[id=0]{throughout training} \added[id=0]{over five training phases}. \added[id=R1]{We found that starting with small polygons and increasing the size and complexity resulted in stable training. We decrease the entropy coefficient as training progresses in order to discourage ``exploration" of new strategies in favor of convergence of the policy. Increasing the clip parameter and learning rate allow for more aggressive updates of the model weights as training plateaus.} To generate a random polygon with $k$ sides, we randomly sample $k$ numbers in $[.6,1]$ and $k-1$ angles that sum to $2 \pi$. These $k$ points in polar coordinates form a polygon contained in the unit circle. We scale the coordinates of this polygon by some number to create a larger domain. We add boundary vertices to the line segments that make up polygonal boundary by splitting each edge into $\max( \hbox{ceil}(|e|/1.4),1)$ segments. See \deleted[id=0]{table} \added[id=0]{Table} \ref{tab:hyper} for hyperparameter settings.

\begin{table}[]
    \centering
    \begin{tabular}{|c|c|c|c|c|c|}
        \hline
        \added[id=R2]{Training Phase} & \added[id=R2]{I} & \added[id=R2]{II} & \added[id=R2]{III} & \added[id=R2]{IV} & \added[id=R2]{V} \\
        \hline
        Polygon scaling & 2 & 2.5 & 3 & 3.5 & 4 \\ \hline
        Num. side range & 5-10 & 5-12 & 5-14 &5-16 & 5-18  \\ \hline
        Trajectory length & 5 & 6 & 7 & 8 & 9 \\ \hline
        Entropy Coef. & 1e-2 & 1e-3 & 1e-4 & 1e-5 & 1e-6 \\ \hline
        Learning rate & 1e-5 & 1.5e-5 & 2e-5 & 2.5e-5 & 3e-5 \\ \hline
        PPO $\epsilon$ & .1 & .15 & .2 & .25 & .3 \\ \hline
        Num. PPO It. & 200 & 200 & 200 & 200 & 200 \\ \hline
        Epoch/It. & 10 & 10 & 10 & 10 & 10 \\ \hline
        Trajectory/It. & 300 & 300 & 300 & 300 & 300 \\ \hline
    \end{tabular}
    \caption{Training curriculum hyperparameters.}
    \label{tab:hyper}
\end{table}

\subsection{Model Size Experiment}
In this experiment we study the effect of varying the size of the neural networks used for the policy and value functions. We vary the number of convolutional layers $\ell$ and the dimension $d$ of the hidden node feature $z_i$. In each convolutional layer, each node exchanges information with its neighbors, so having $\ell$ convolutional layers means a node can communicate with its entire $\ell$-hop neighborhood. Although we leave most computational concerns for future research, by using a graph neural network we have ensured good scaling: for a mesh with $N$ vertices, the computational cost of a convolutional layer scales like a sparse $N \times N$ matrix times a dense $N \times d$ matrix, i.e., linearly in the number of vertices. In addition, this operation is amenable to the same sort of spatial parallelization that is used in finite element codes. For large meshes the cost of the Delaunay triangulation algorithm will be greater than running the policy function, which could be improved by incorporating local Delaunay updates instead of complete retriangulations.

\begin{figure}
    \centering
\includegraphics[width=1\linewidth,height=8cm,keepaspectratio]{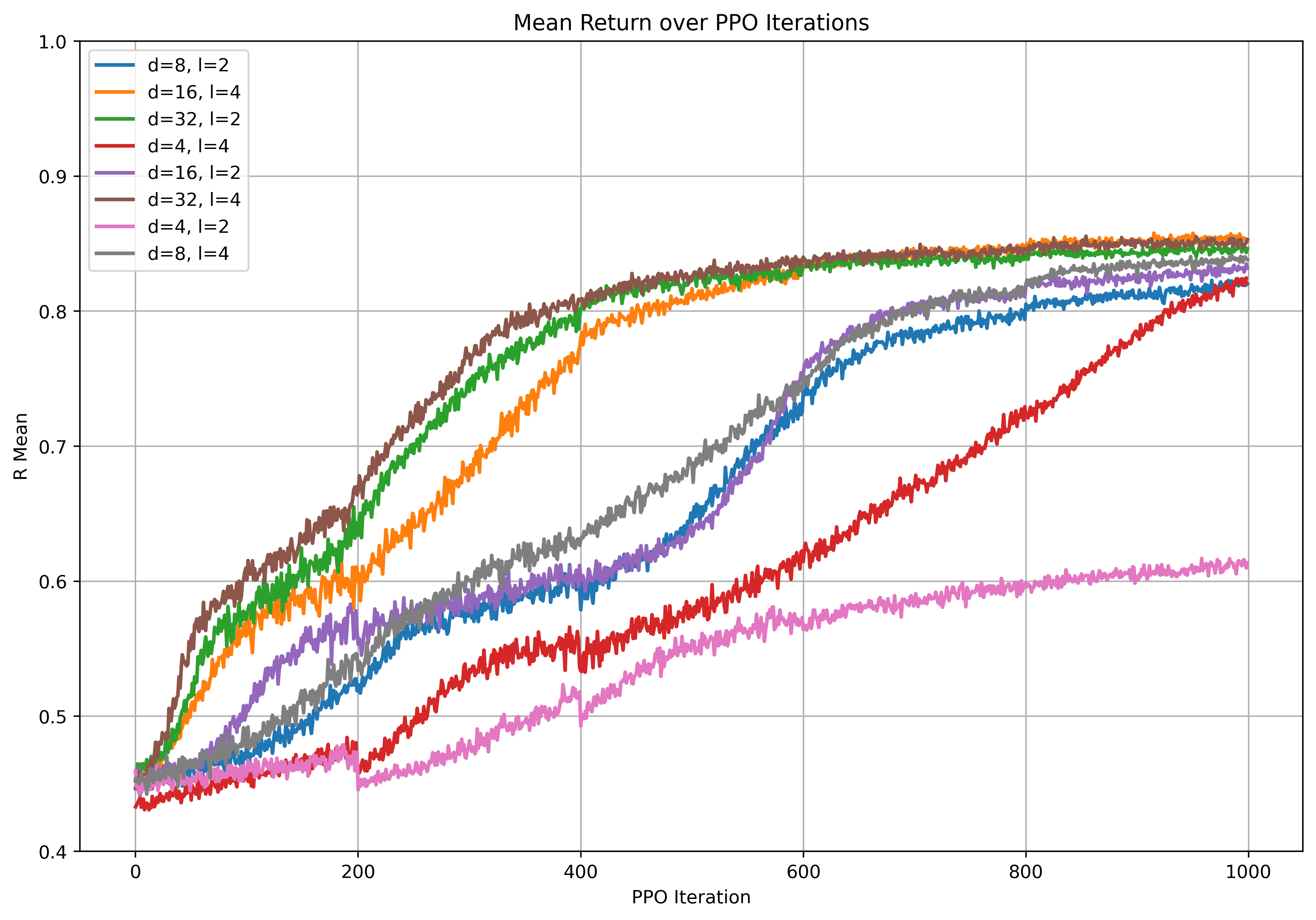}
    \caption{Mean reward over training for different model sizes. For this training curriculum, we sample 2.1 million timesteps over 1000 PPO iterations.}
    \label{fig:training-size}
\end{figure}

The results of the experiment do not strongly indicate that increasing model size improves performance. Training curves are shown in Figure \ref{fig:training-size}. Reported mesh metrics are for deterministic policies, which we found to perform better for this particular experiment. The two top performing models are $\ell=2, d=8$ and $\ell=4, d = 16$, and both achieve an average reward, defined in \eqref{eq:rew}, of about $.86$ for the final training curriculum settings. Note that the model that achieves the highest training reward does not perform the best on the evaluation. This experiment is not perfect; it may be that better performance is achievable with a different training curriculum or more detailed hyperparameter optimization. We use the $\ell=2, d=8$ model as the baseline model in the remainder of our experiments, as the $\ell=4, d = 16$ model has about seven times as many parameters. We experimented with additional training curriculum rounds, but the evaluation scores do not improve.

\begin{figure}
    \centering
    \subfloat[50]{\includegraphics[trim={70 20 80 30},clip,width=.5\linewidth,height=5cm,keepaspectratio]{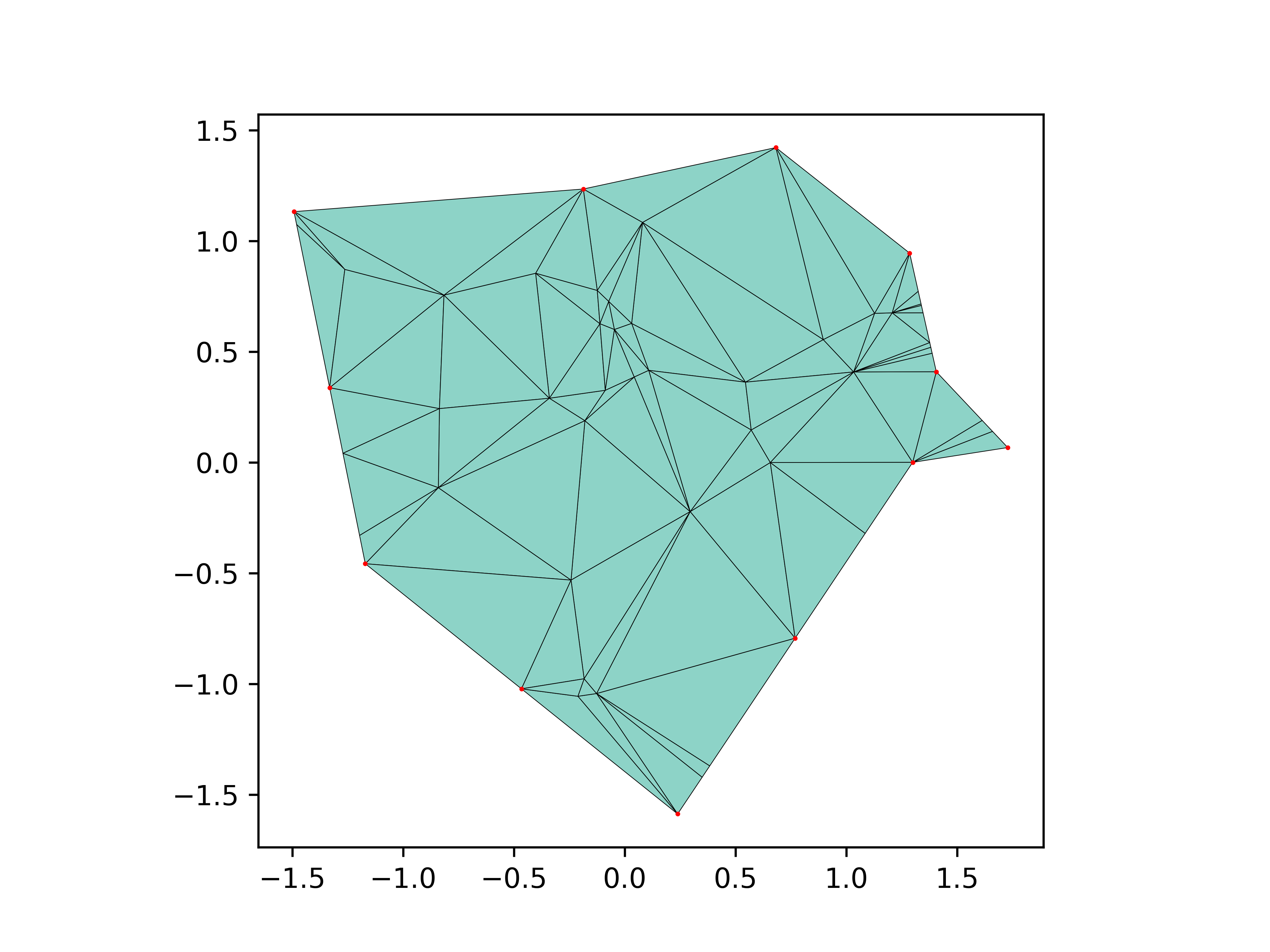}}
    \subfloat[250]{\includegraphics[trim={70 20 80 30},clip,width=.5\linewidth,height=5cm,keepaspectratio]{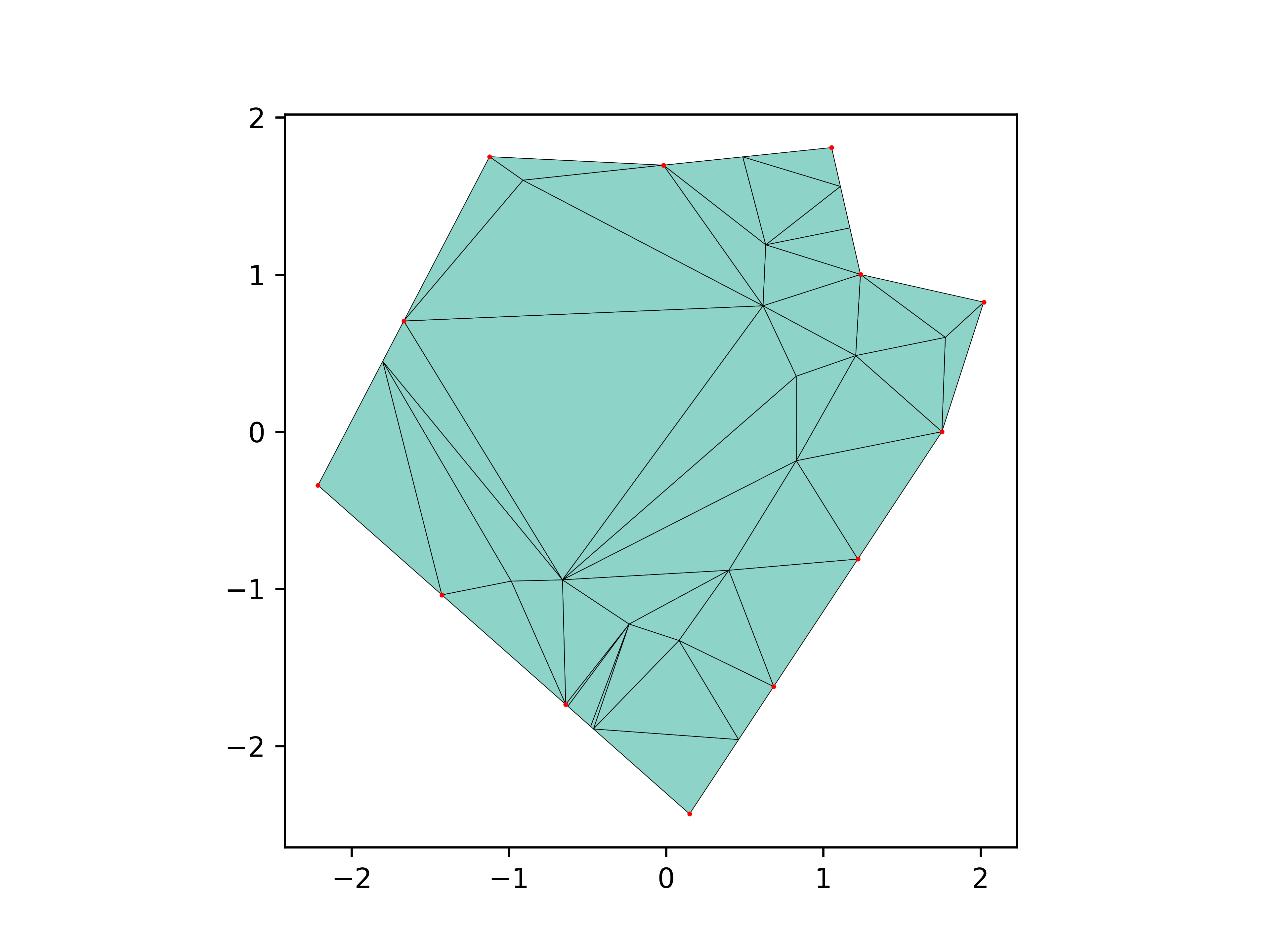}}
    \hspace{0mm}
    \subfloat[450]{\includegraphics[trim={70 20 80 30},clip,width=.5\linewidth,height=5cm,keepaspectratio]{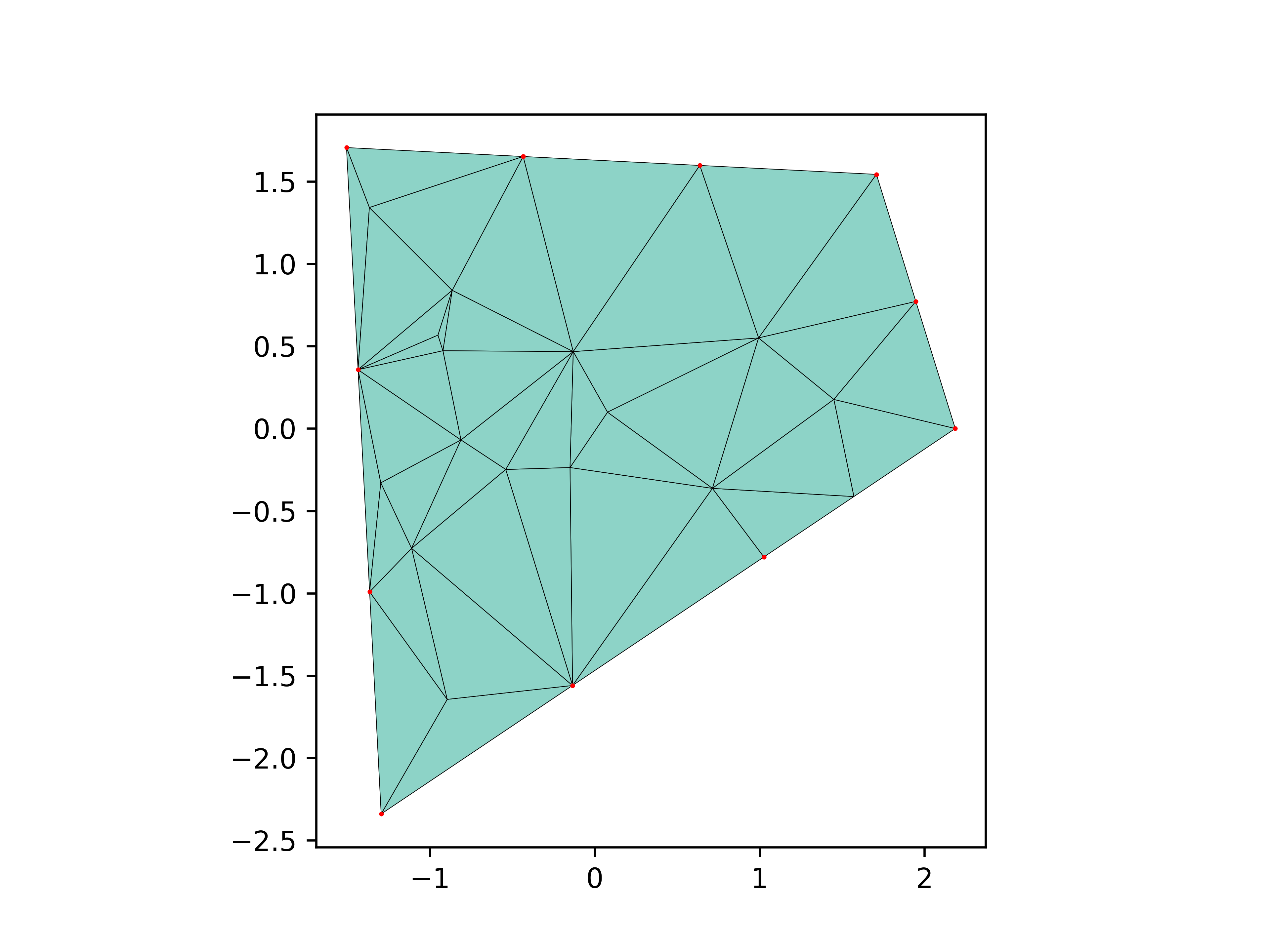}}
    \subfloat[650]{\includegraphics[trim={70 20 80 30},clip,width=.5\linewidth,height=5cm,keepaspectratio]{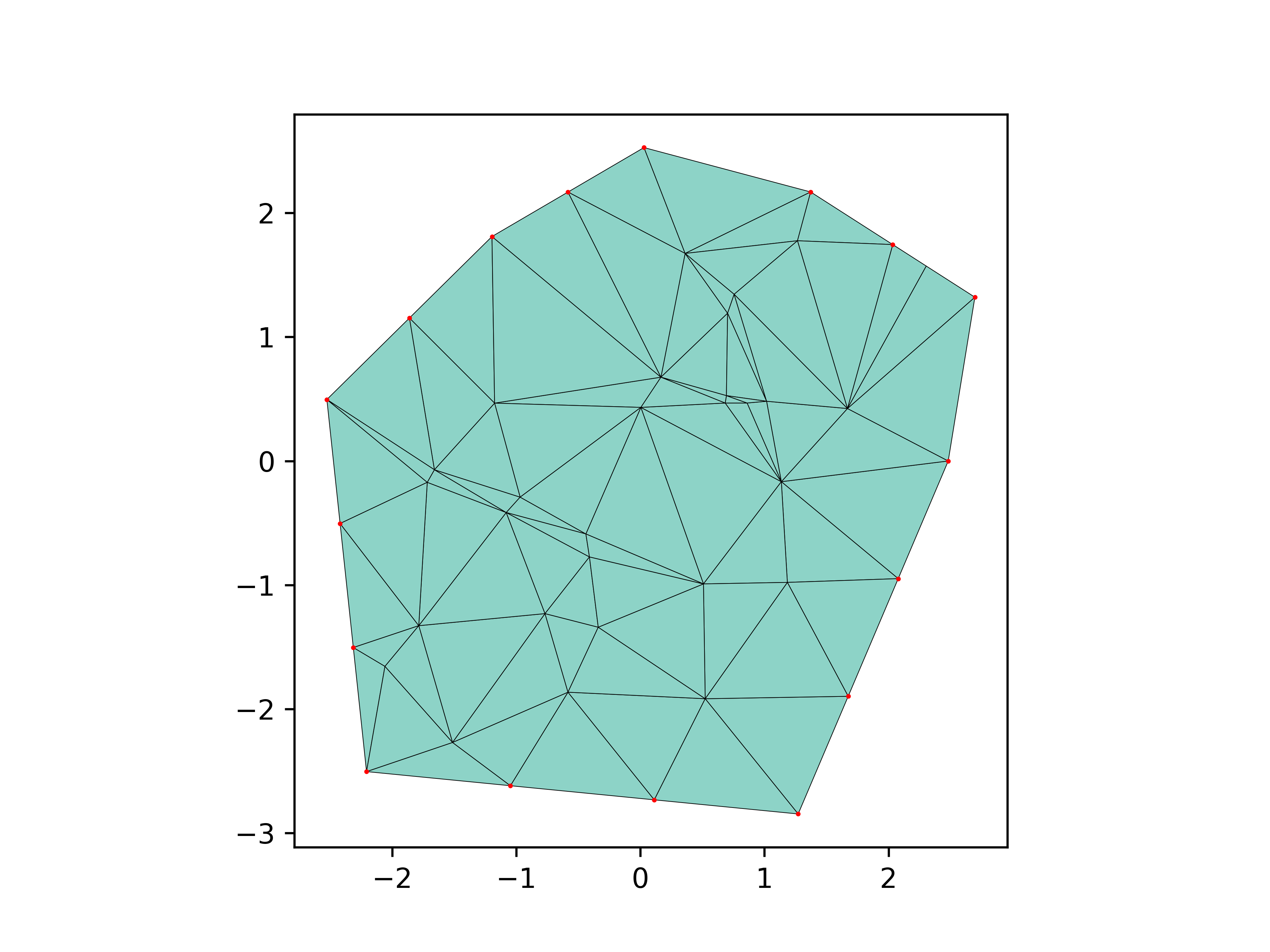}}
    \hspace{0mm}
    \subfloat[850]{\includegraphics[trim={70 20 80 30},clip,width=.5\linewidth,height=5cm,keepaspectratio]{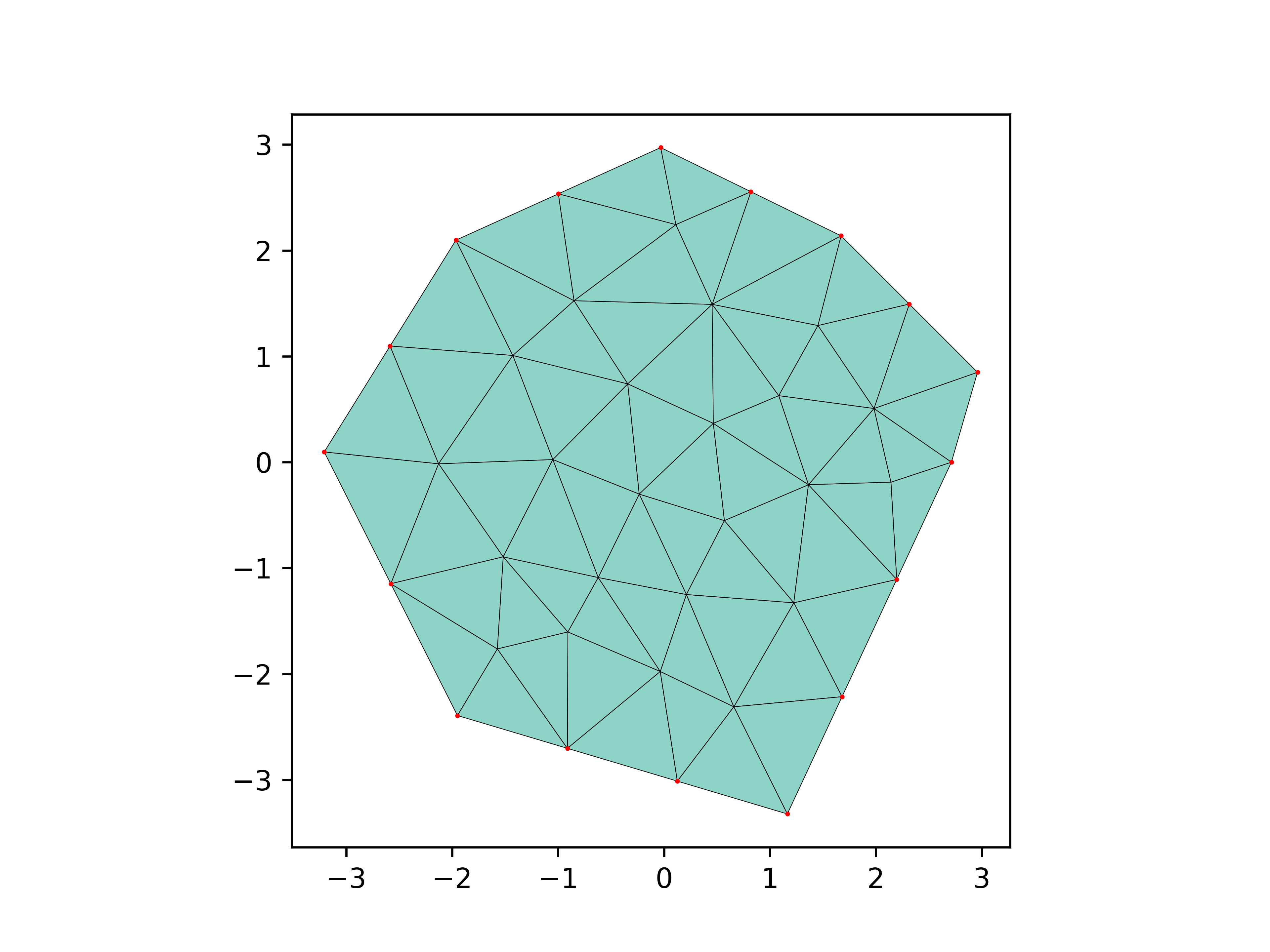}}
    \subfloat[1000]{\includegraphics[trim={70 20 80 30},clip,width=.5\linewidth,height=5cm,keepaspectratio]{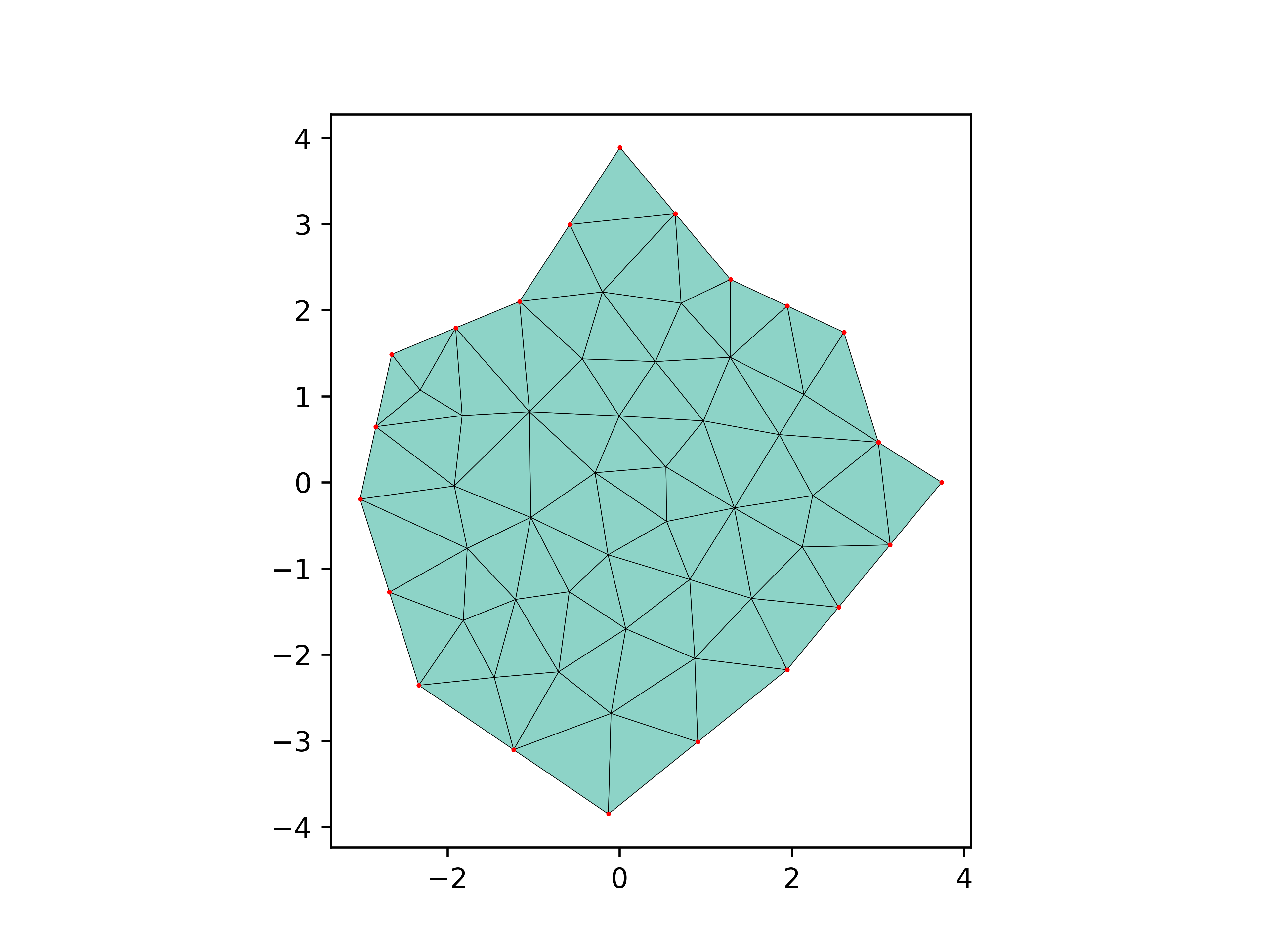}}
    \hspace{0mm}
    \caption{Examples of meshes produced during training at PPO iteration 50, 250, 450, 650, 850 and 1000, ordered from top left to bottom right. The target edge length is 1, indicated by the scaling on the axes.}
    \label{fig:training-meshes}
\end{figure}

\begin{figure}
    \centering
    \subfloat{\includegraphics[trim={850 400 750 450},clip,width=.5\linewidth,height=4cm,keepaspectratio]{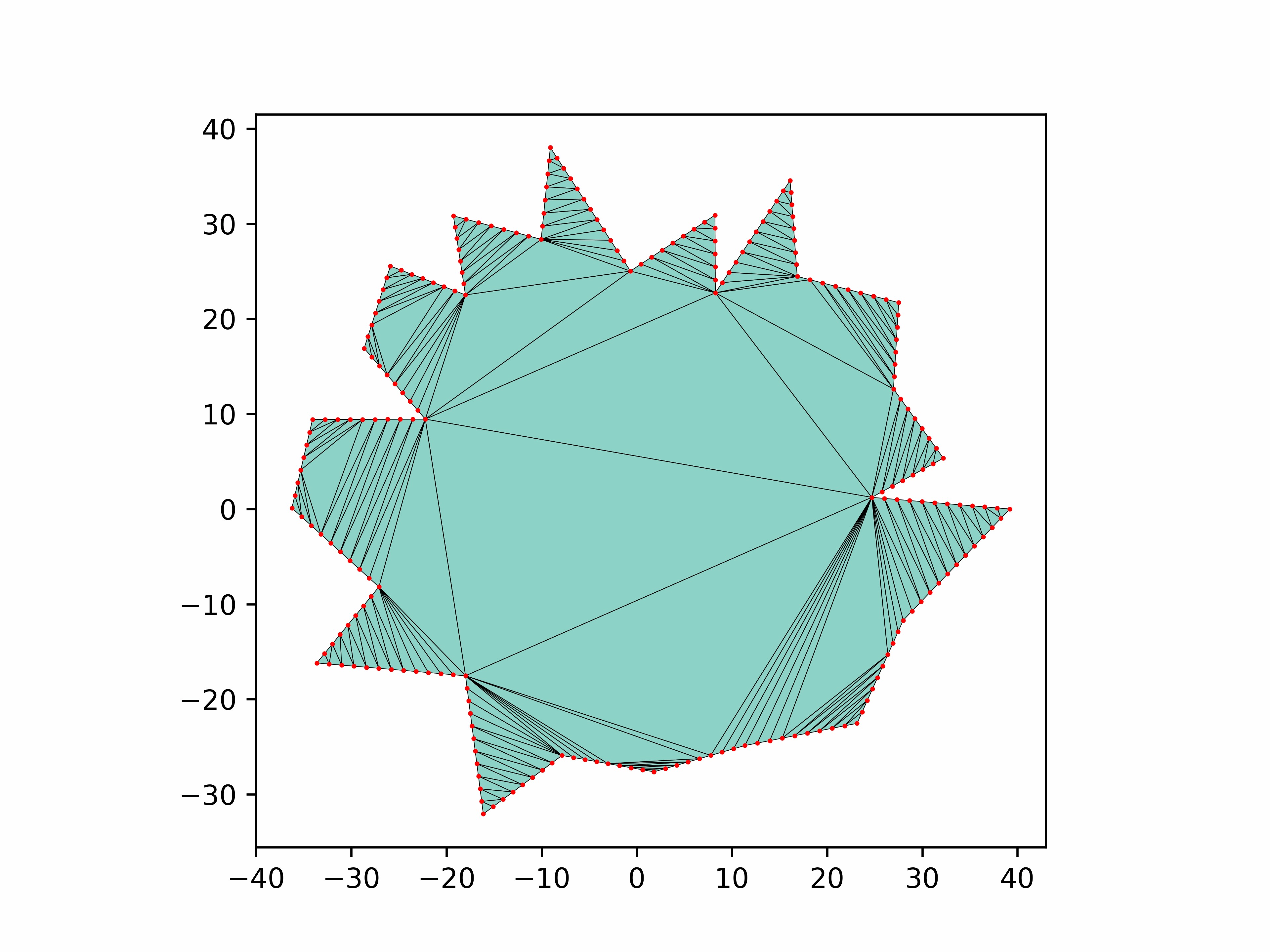}}
    \subfloat{\includegraphics[trim={850 400 750 450},clip,width=.5\linewidth,height=4cm,keepaspectratio]{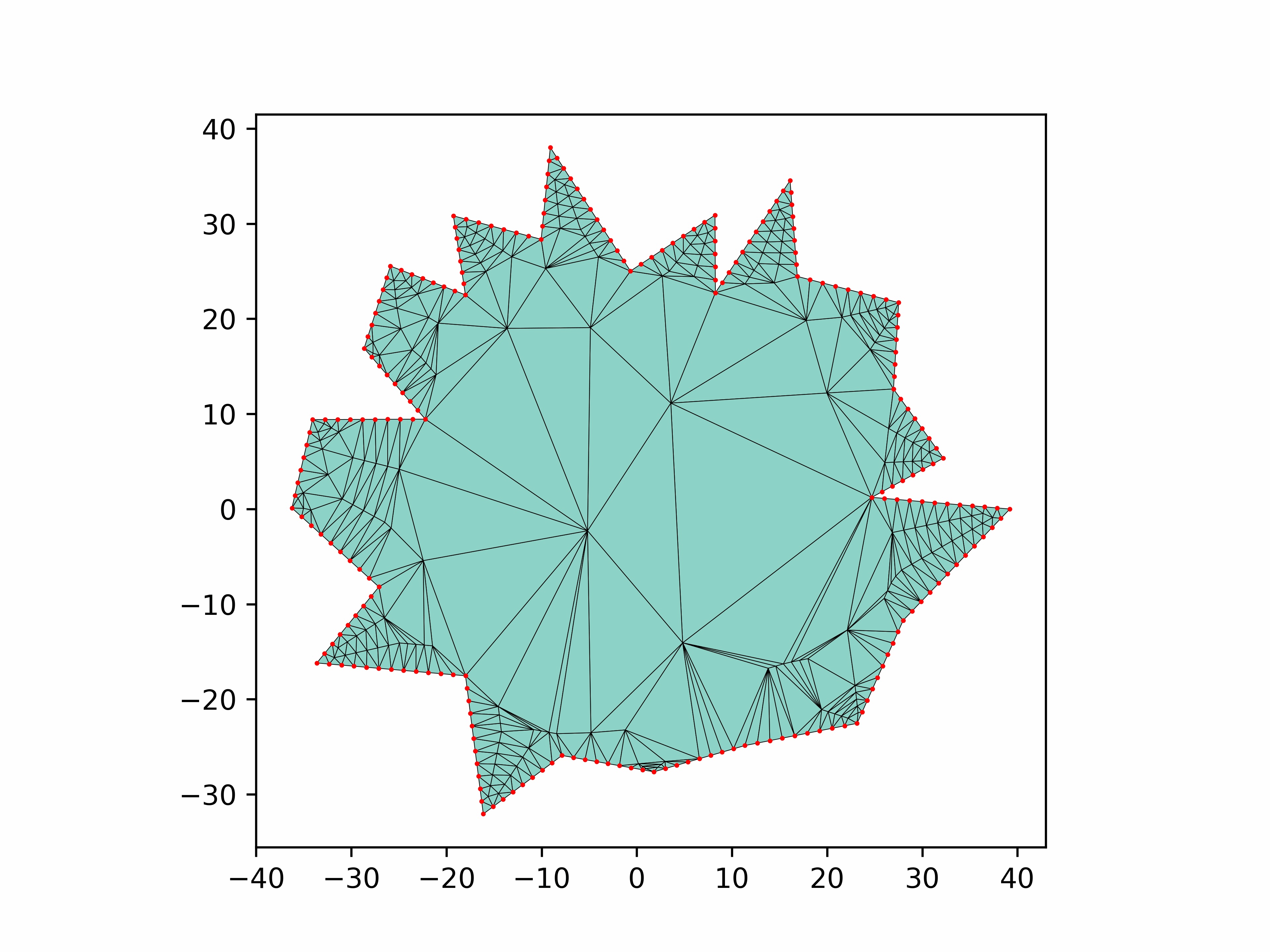}}
    \hspace{0mm}
    \subfloat{\includegraphics[trim={850 400 750 450},clip,width=.5\linewidth,height=4cm,keepaspectratio]{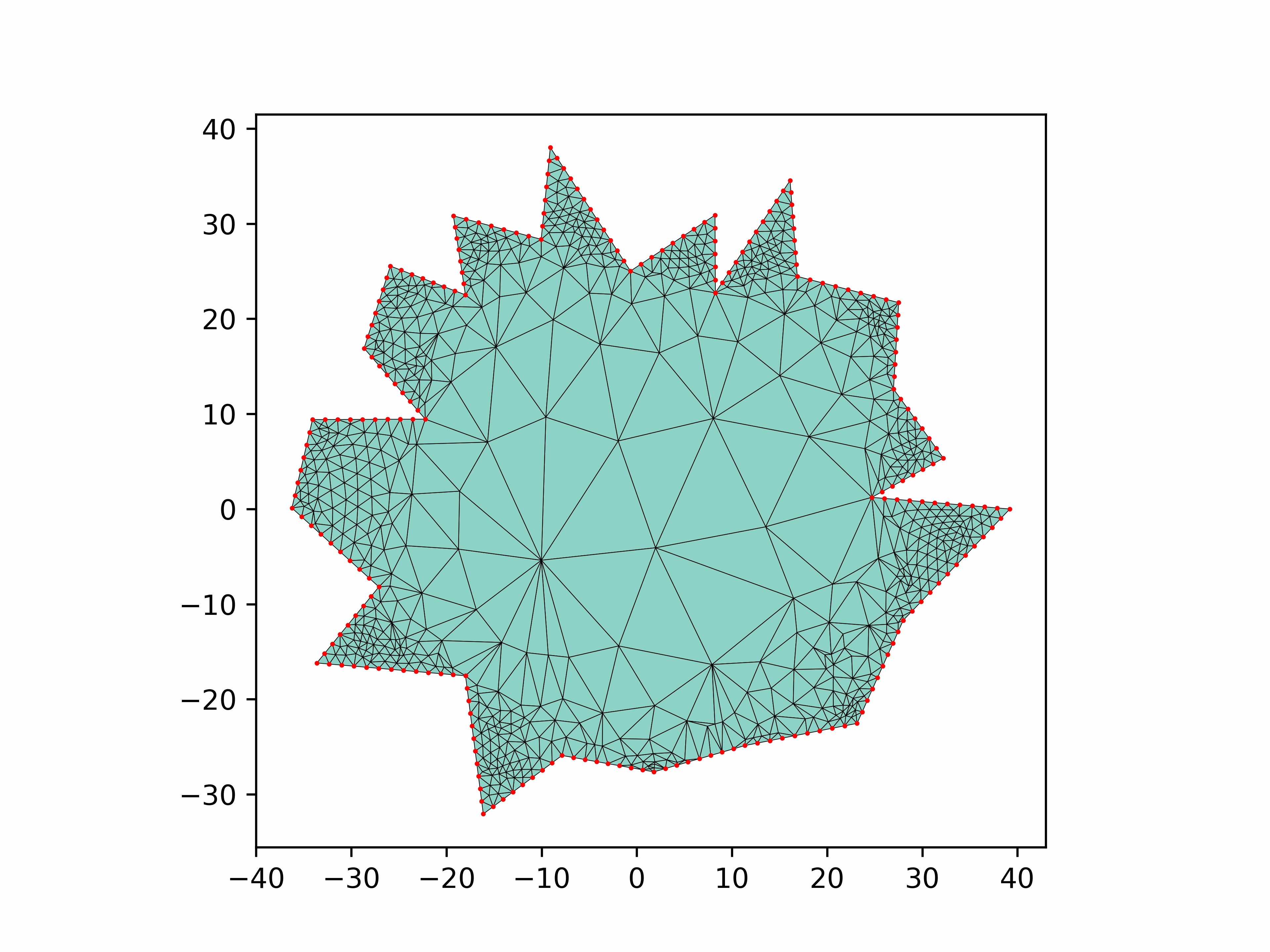}}
    \subfloat{\includegraphics[trim={850 400 750 450},clip,width=.5\linewidth,height=4cm,keepaspectratio]{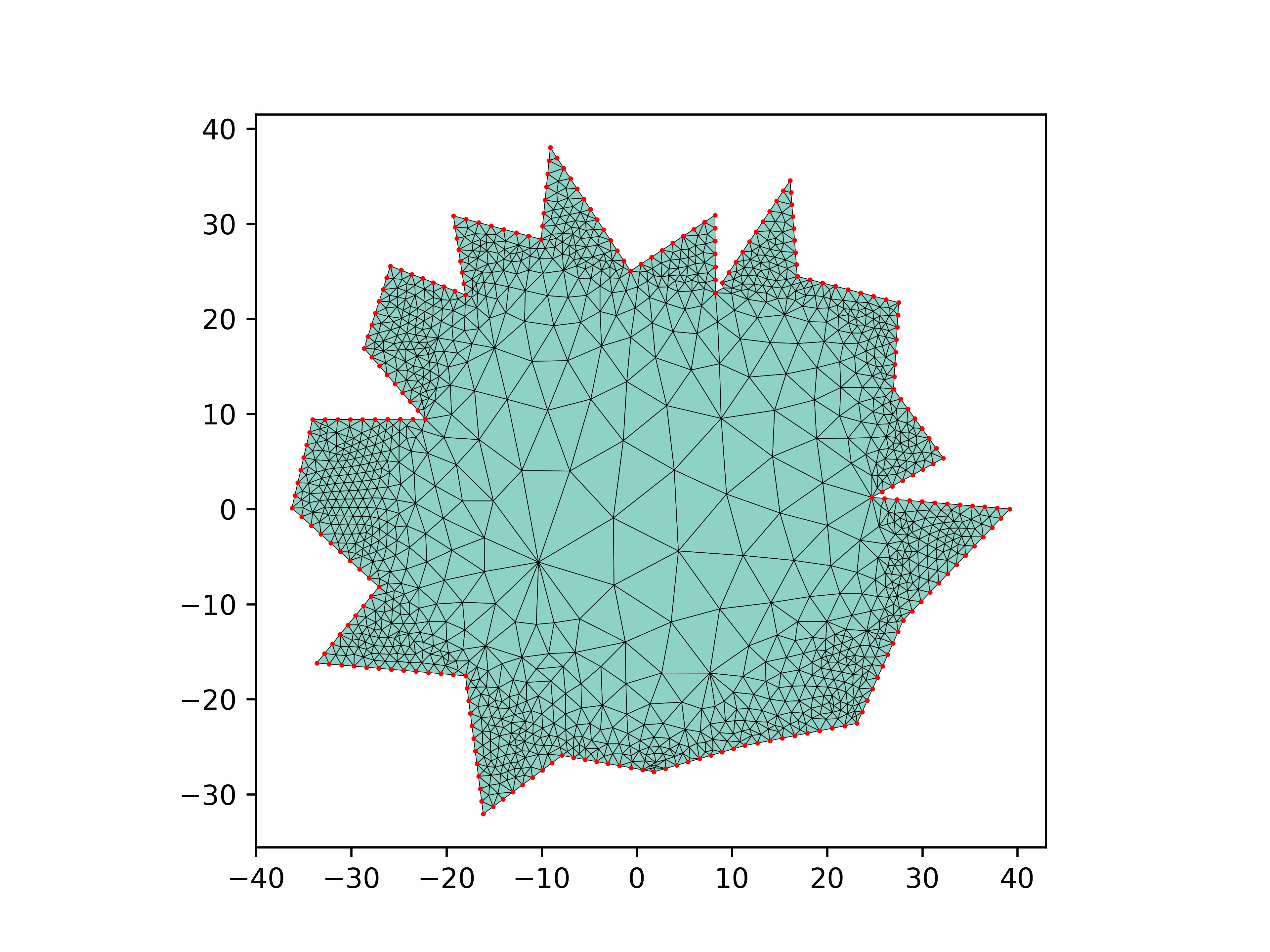}}
    \hspace{0mm}
    \subfloat{\includegraphics[trim={850 400 750 450},clip,width=.5\linewidth,height=4cm,keepaspectratio]{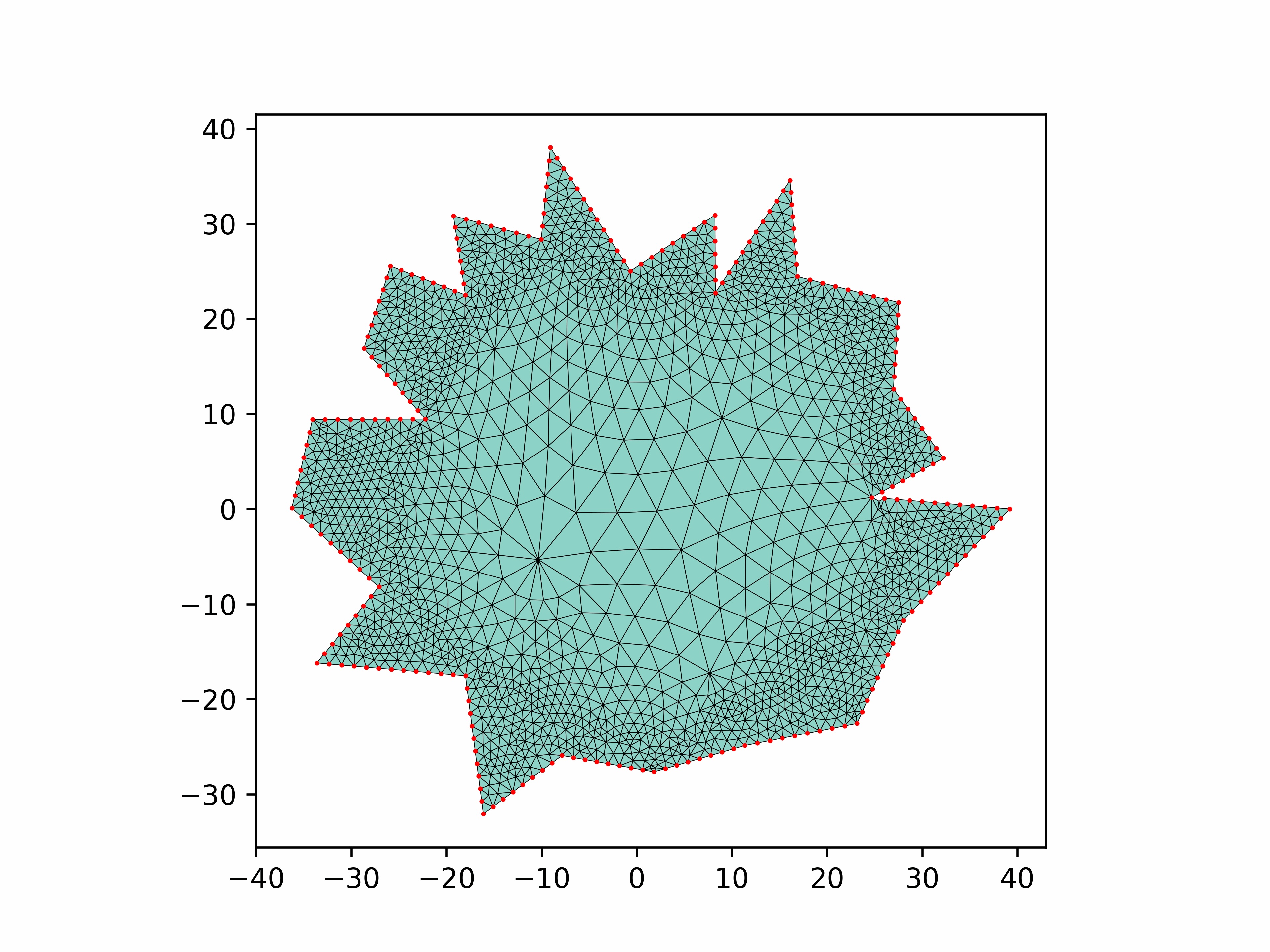}}
    \subfloat{\includegraphics[trim={850 400 750 450},clip,width=.5\linewidth,height=4cm,keepaspectratio]{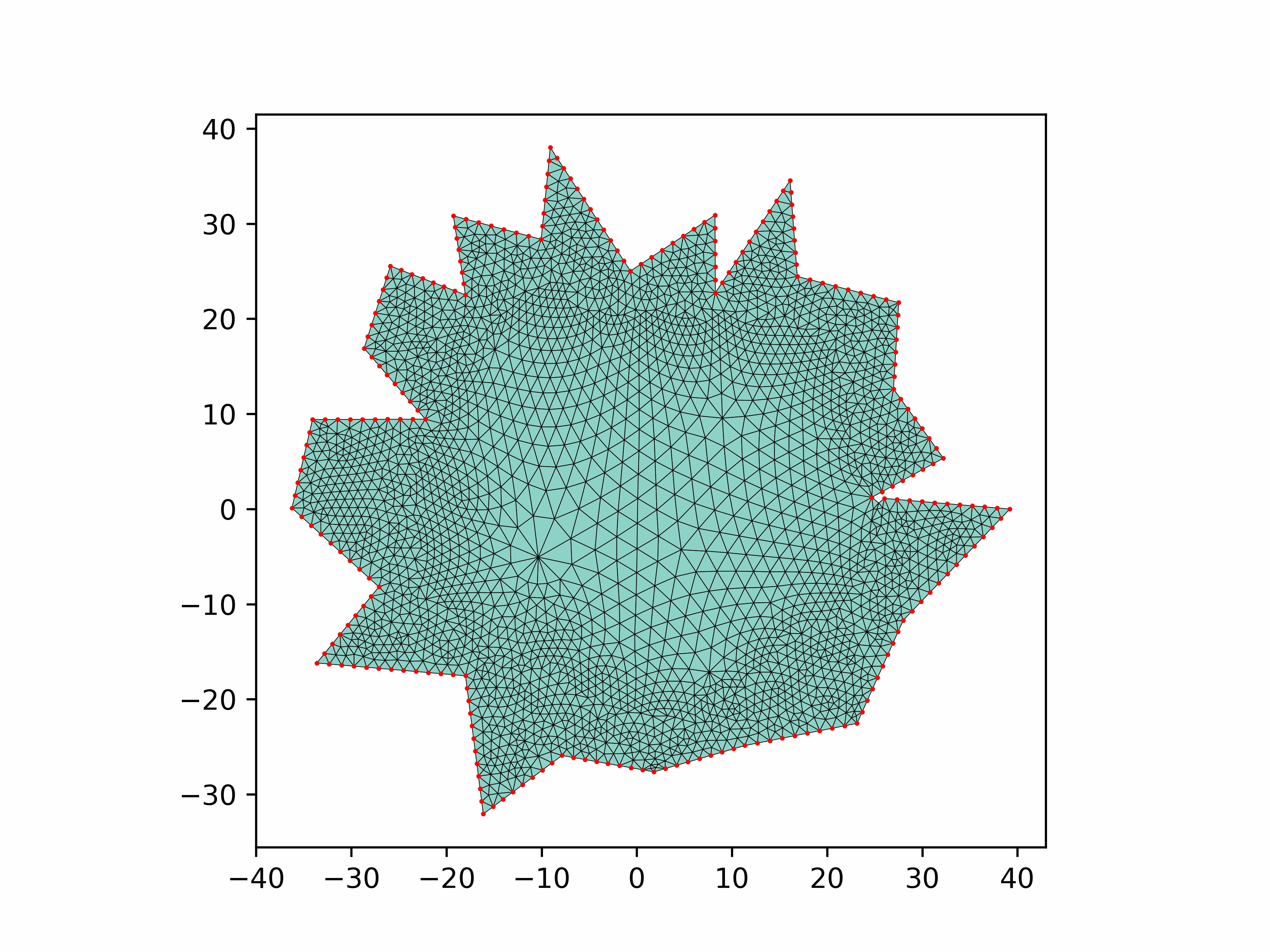}}
    \hspace{0mm}
    \subfloat{\includegraphics[trim={850 400 750 450},clip,width=.5\linewidth,height=4cm,keepaspectratio]{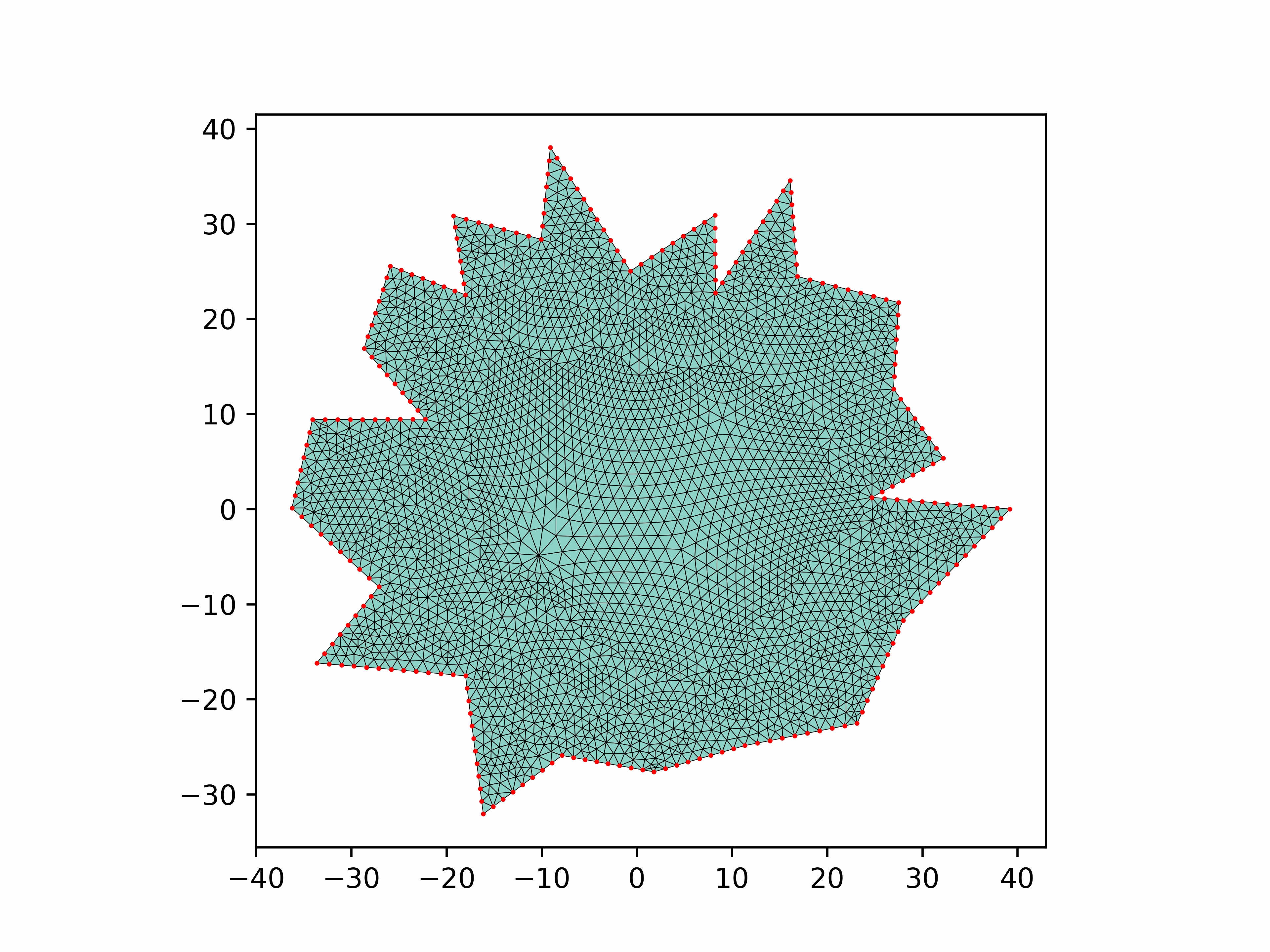}}
    \subfloat{\includegraphics[trim={850 400 750 450},clip,width=.5\linewidth,height=4cm,keepaspectratio]{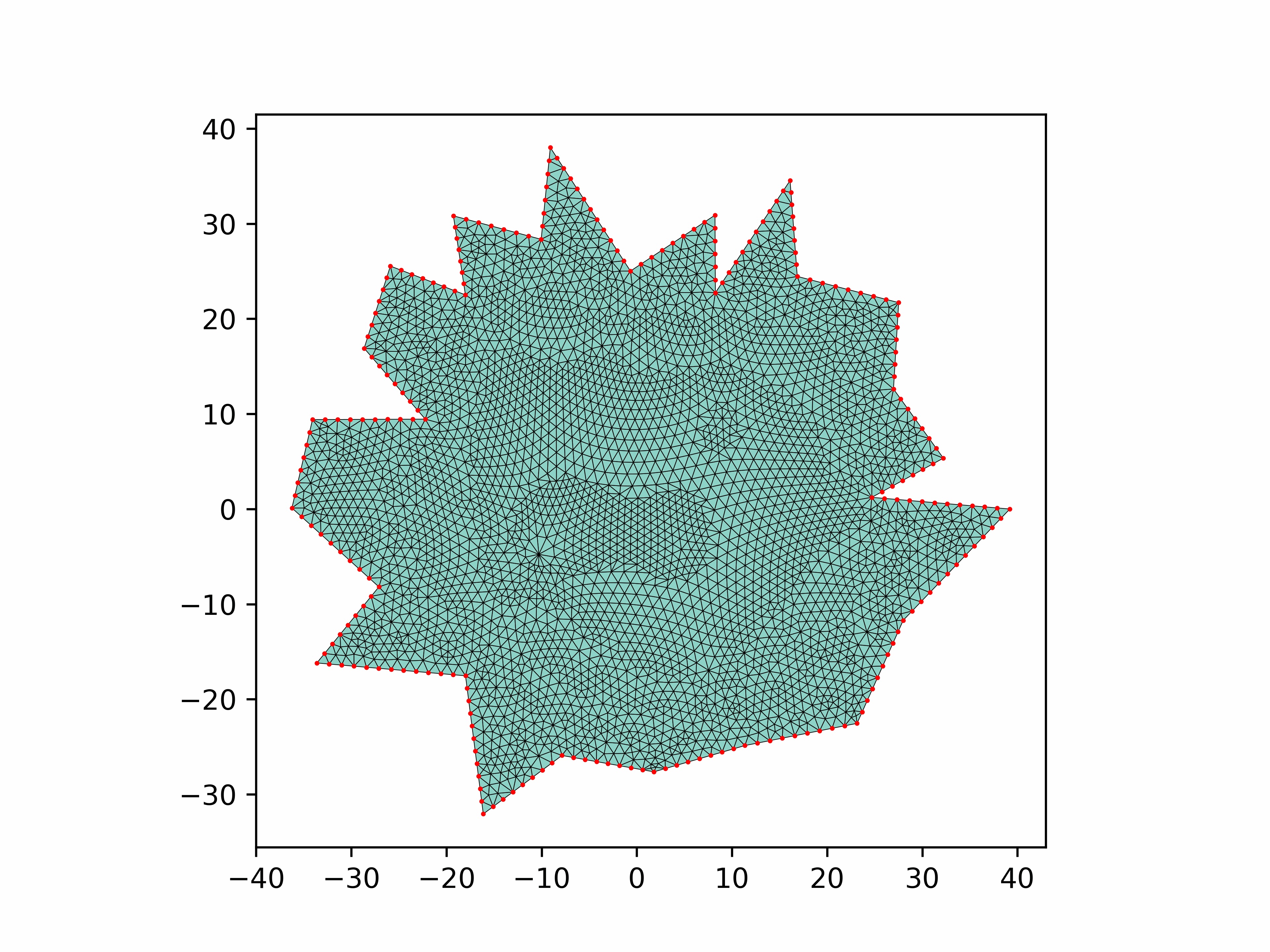}}
    \hspace{0mm}
    \caption{Trained mesh generator on a polygon with scaling factor of 40. Convergence to the final mesh is achieved in about 10 iterations. }
    \label{fig:mesh-gif}
\end{figure}

Figure \ref{fig:training-meshes} shows meshes produced at subsequent stages of training. In Figure \ref{fig:mesh-gif} we see the fully trained algorithm meshing a polygon that is scaled by a factor of 40, with a final mesh that has on the order of $\sim10^3$ elements. We observe that the vertex positions converge and no new vertices are added. Even for larger meshes, the algorithm only takes a small number of steps to converge to a final mesh because the number of vertices can more than double each timestep.


\begin{table}[h!]
\centering
\resizebox{\textwidth}{!}{\begin{tabular}{|c|c|c|c|c|c|c|c|c|c|c|c|c|c|c|}
\hline 
d & $\ell$ & $|\theta|$ & $q_a$ (Mean) & $q_a$ (Min) & $q_a$ (SD) & $q_e$ (Mean) & $q_e$ (Min) & $q_e$ (SD) & $q_r$ (Mean) & $q_r$ (Min) & $q_r$ (SD) & $q_v$ (Mean) & $q_v$ (Min) & $q_v$ (SD) \\ \hline
4 & 2 & 1107 & 0.841  & 0.111  & 0.133  & 0.582  & -0.252  & 0.256  & 0.936  & 0.582  & 0.070  & 0.089 &-1.976  & 0.632     \\ \hline
8 & 2 & 3791 & \textbf{0.860}  & \textbf{0.198}  & 0.118  & 0.862  & 0.327  & 0.114  & \textbf{0.948}  & \textbf{0.619}  & 0.060  & 0.779 &-0.127  & 0.182   \\ \hline
16 & 2 & 13959& 0.821  & 0.100  & 0.136  & 0.857  & 0.359  & 0.110  & 0.922  & 0.552  & 0.073  & 0.795 &-0.067  & 0.161   \\ \hline
32 & 2 & 53495 & 0.839  & 0.180  & 0.123  & 0.877  & 0.497  & 0.092  & 0.936  & 0.590  & 0.062  & 0.857 &0.293  & 0.119   \\ \hline
4 & 4 & 2029& 0.836  & 0.102  & 0.131  & 0.853  & 0.404  & 0.111  & 0.932  & 0.544  & 0.071  & 0.765 &0.079  & 0.1599   \\ \hline
8 & 4 & 7033 & 0.824  & 0.041  & 0.136  & 0.873  & 0.425  & 0.098  & 0.924  & 0.494  & 0.075  & 0.821 &0.338  & 0.130   \\ \hline
16 & 4 & 26065 & 0.839  & 0.191  & 0.123  & \textbf{0.883}  & \textbf{0.519}  & 0.089  & 0.936  & 0.597  & 0.062  & \textbf{0.865} & \textbf{0.388}  & 0.108    \\ \hline
32 & 4 & 100225 & 0.826  & 0.101  & 0.133  & 0.873  & 0.366  & 0.102  & 0.926  & 0.556  & 0.070  & 0.842 &0.028  & 0.144    \\ \hline
\end{tabular}}
\caption{Mesh quality metrics for model size test.}
\label{tab:stat_summary}
\end{table}

\subsection{Modification of reward function} \label{sec:mod}
In this section we demonstrate that we can optimize for different mesh quality metrics. We use reward functions which are weighted averages of the form $w_a q_a + w_e q_e + w_r q_r + w_v q_v$, where each quality metric is averaged over the mesh. For each reward function, we train a model using the same hyperparameters and curriculum as the prior experiment. For the previous experiment, we used a deterministic policy, meaning we manually set the variance to zero so that the mean action was always sampled. However, \added[id=R1]{it is common to use a stochastic policy at inference time when the policy is trained using PPO \cite{schulman2017proximalpolicyoptimizationalgorithms}, and} in this section we also report results on stochastic policies, meaning we do not modify the variance \added[id=R1]{outputted by the policy network}.  Results are shown in Table \ref{tab:rew-vary}.

These results indicate that we can improve a particular mesh metric by specifically training for it. If we only include edges or only include volumes in our reward, we improve the score in that category relative to the base model. While we did experiment with increasing the weight on the angle quality, we were not able to improve upon the baseline model. For the model trained solely on the volume reward, as well as the one trained on the evenly balanced reward, the stochastic policy obtains better results than the deterministic policy.  There are many potential explanations for this including hyperparameter settings, the training curriculum, the optimization algorithm, or the policy network architecture. \added[id=R1]{This phenomenon may have complex causes, and understanding it in greater detail is beyond the scope of this paper and will be the subject of future research. For the purposes of this work we emphasize that we were able to train different policies to maximize different mesh metrics by changing the reward function; the fact that some of these policies turned out to be stochastic is not unusual in the context of reinforcement learning.} \deleted[id=R1]{In future work we will attempt to understand this phenomenon in greater detail.}

\begin{figure}
    \centering
    \subfloat{{\includegraphics[trim={100 300 100 400},clip,width=0.5\linewidth]{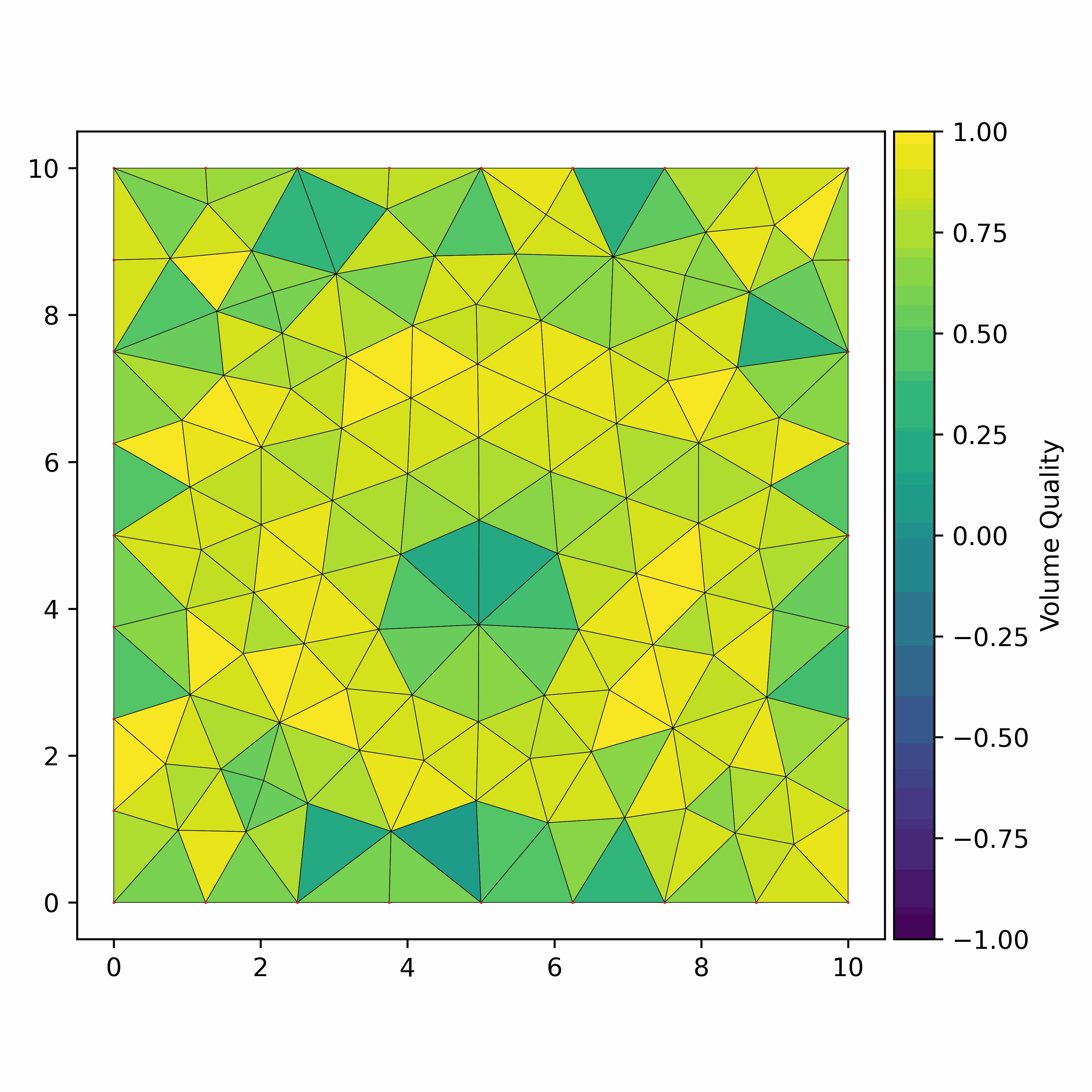}}}
    \subfloat{\includegraphics[trim={100 300 100 400},clip,width=0.5\linewidth]{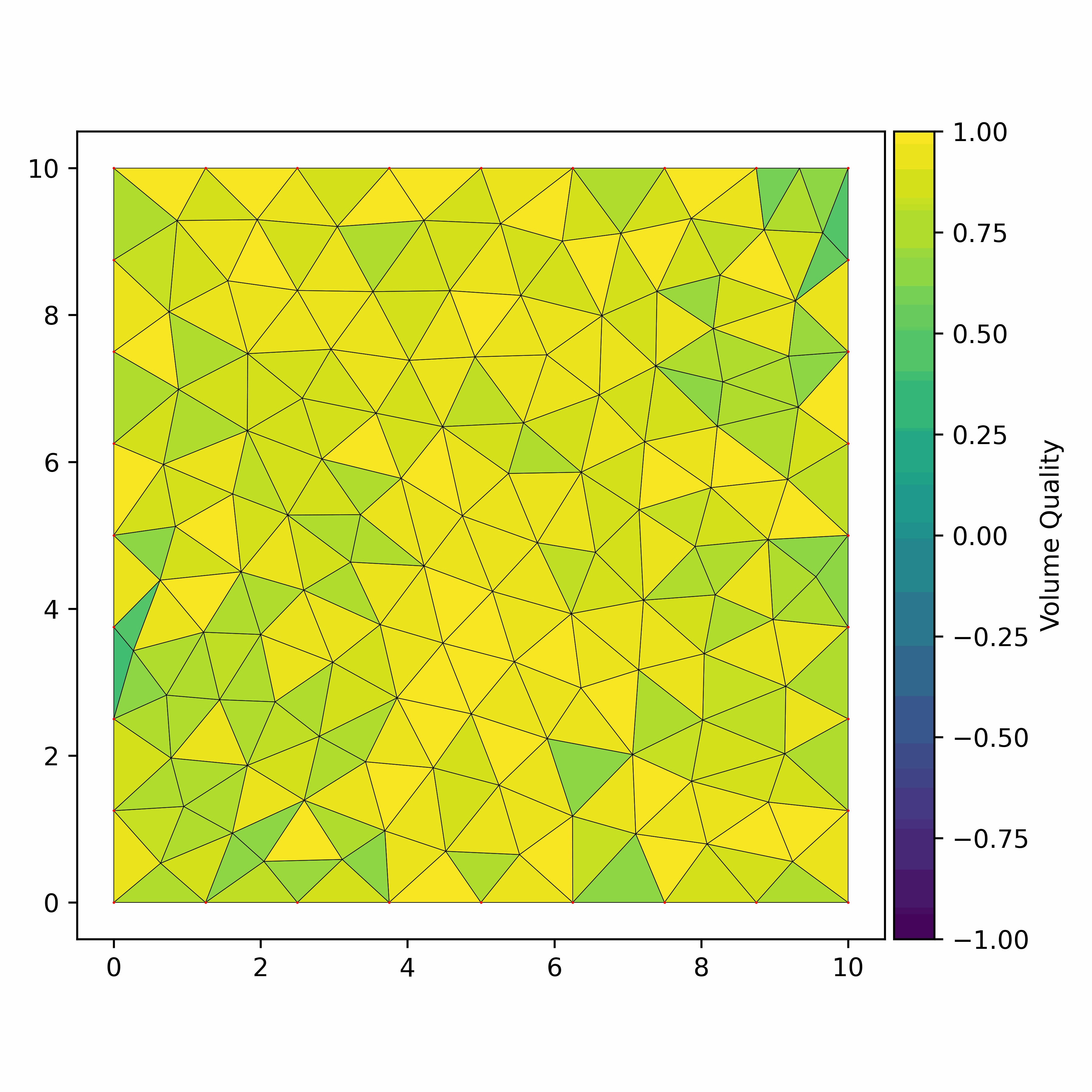}}
    \caption{Meshes produced from policies trained with different reward functions: on the left we trained using $w_a=w_e=.5$ and use a deterministic policy and on the right we trained using $w_v=1$ and use a stochastic policy.}
    \label{fig:rew-vary}
\end{figure}

A visualization of the difference in meshes induced by training with different score metrics is shown in Figure \ref{fig:rew-vary}. We display element volumes to show the improvement that can be gained over the baseline model. It appears that these policies have made certain trade-offs in their learned decision making processes; elements in the left figure are generally better shaped and more uniform but are sometimes too large, while those on the right are not always well-shaped but are more uniformly of the target volume. While we only highlight a few examples of different score functions here, this indicates that our mesh generator framework is indeed flexible enough to be adapted for different desired mesh outcomes.
\begin{table}[h!]
\centering
\resizebox{\textwidth}{!}{\begin{tabular}{|c|c|c|c|c|c|c|c|c|c|c|c|c|c|c|c|c|}
\hline
 Det. & $w_a$ & $w_r$ & $w_e$ & $w_v$ &$q_a$ (Mean) & $q_a$ (Min) & $q_a$ (SD) & $q_e$ (Mean) & $q_e$ (Min) & $q_e$ (SD) & $q_r$ (Mean) & $q_r$ (Min) & $q_r$ (SD) & $q_v$ (Mean) & $q_v$ (Min) & $q_v$ (SD) \\ \hline
Y &.5 & 0 & .5 & 0 & \textbf{0.860}  & 0.198  & 0.118  & 0.862  & 0.327  & 0.114  & \textbf{0.948}  & \textbf{0.619}  & 0.060  & 0.779 &-0.127  & 0.182   \\ \hline
N &.5 & 0 & .5 & 0 & 0.841  & 0.043  & 0.129  & 0.847  & 0.378  & 0.113  & 0.936  & 0.484  & 0.070  & 0.751 &0.137  & 0.170  \\ \hline
Y & 0 & 0 & 1 & 0 & 0.857  & \textbf{0.202}  & 0.121  & \textbf{0.868}  & \textbf{0.394}  & 0.106  & 0.945  & 0.597  & 0.062  & 0.794 &0.101  & 0.158 \\ \hline
N & 0 & 0 & 1 & 0 & 0.846  & 0.045  & 0.127  & 0.855  & 0.382  & 0.112  & 0.939  & 0.485  & 0.069  & 0.767 &0.111  & 0.167 \\ \hline
Y &0 & 0 & 0 & 1 & 0.827  & 0.140  & 0.132  & 0.765  & 0.082  & 0.166  & 0.928  & 0.605  & 0.068  & 0.588 &-0.841  & 0.291 \\ \hline
N &0 & 0 & 0 & 1 & 0.827  & -0.226  & 0.139  & 0.877  & 0.331  & 0.097  & 0.925  & 0.261  & 0.085  & \textbf{0.864} & \textbf{0.225}  & 0.116 \\ \hline
Y &.25 & .25 & .25 & .25 & 0.823  & 0.162  & 0.134  & 0.690  & 0.063  & 0.195  & 0.924  & 0.609  & 0.070  & 0.401 &-0.897  & 0.350 \\ \hline
N &.25 & .25 & .25 & .25 & 0.836  & -0.020  & 0.129  & 0.846  & 0.367  & 0.115  & 0.934  & 0.451  & 0.070  & 0.776 &-0.012  & 0.173 \\ \hline
\end{tabular}}
\caption{Mesh quality metrics for reward function test. The ``Det." label indicates if the policy is deterministic.}
\label{tab:rew-vary}
\end{table}

\subsection{Incorporation of size function}
In this section we demonstrate that our mesh generator can be used to produce variable resolution meshes according to a size function $h(x)$, which indicates the desired edge length at the point $x$. During training, the size function is constant at 1, meaning the mesh generator is not trained to produce variable meshes. However, our specific choice of convolutional layer allows our mesh generator to generalize to this task: in the update function \eqref{eq:dist_update}, we divide the internode distance by the value of the size function at the midpoint of the edge between nodes. We experimented with various vertex initialization techniques, and the best performing method was to use no internal vertices initially. We use a deterministic policy for this task.

\begin{figure}
    \centering
    \subfloat{{\includegraphics[trim={50 1100 40 1200},clip,width=.5\linewidth,height=6.5cm,keepaspectratio]{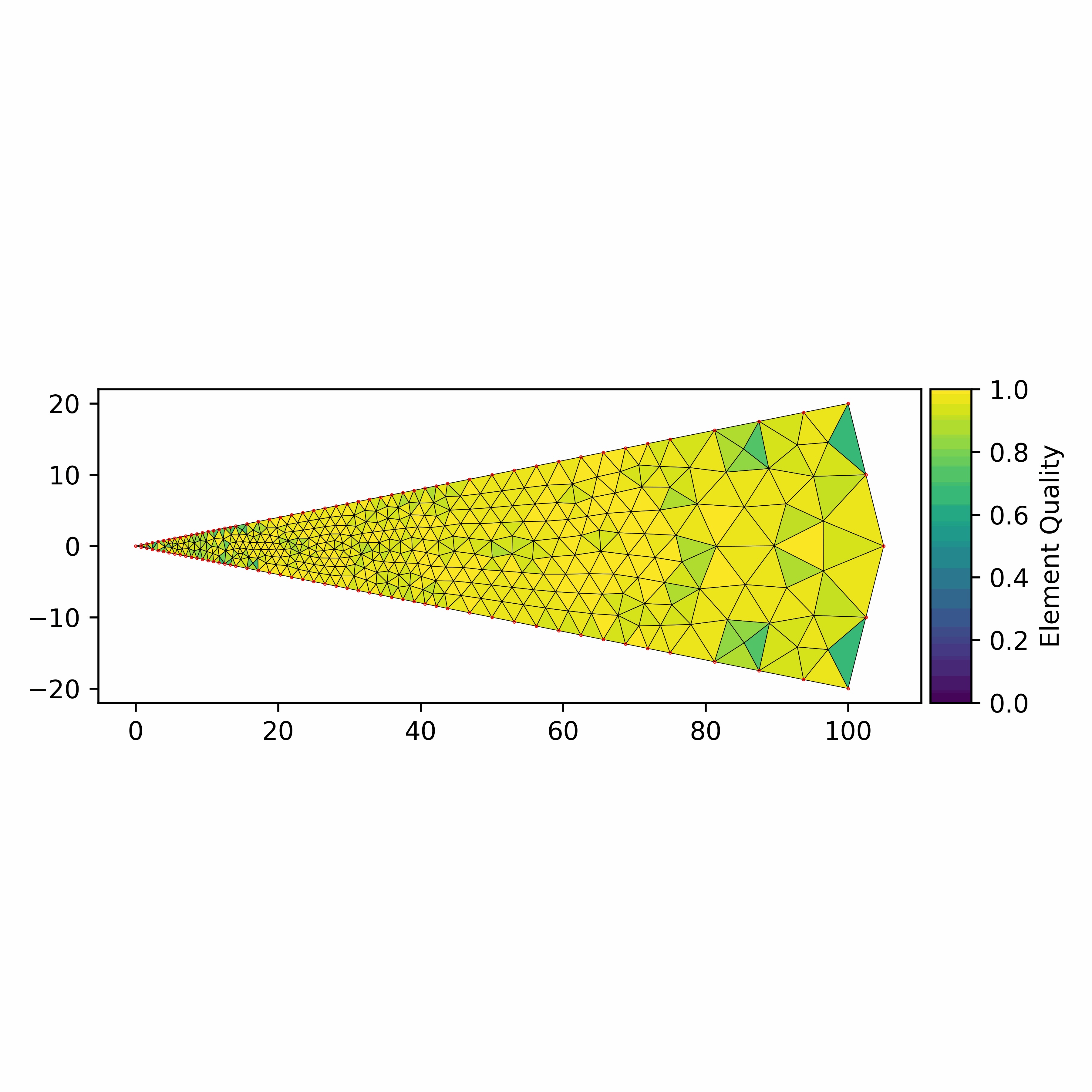}}}
    \subfloat{{\includegraphics[trim={50 1100 40 1200},clip,width=.5\linewidth,height=6.5cm,keepaspectratio]{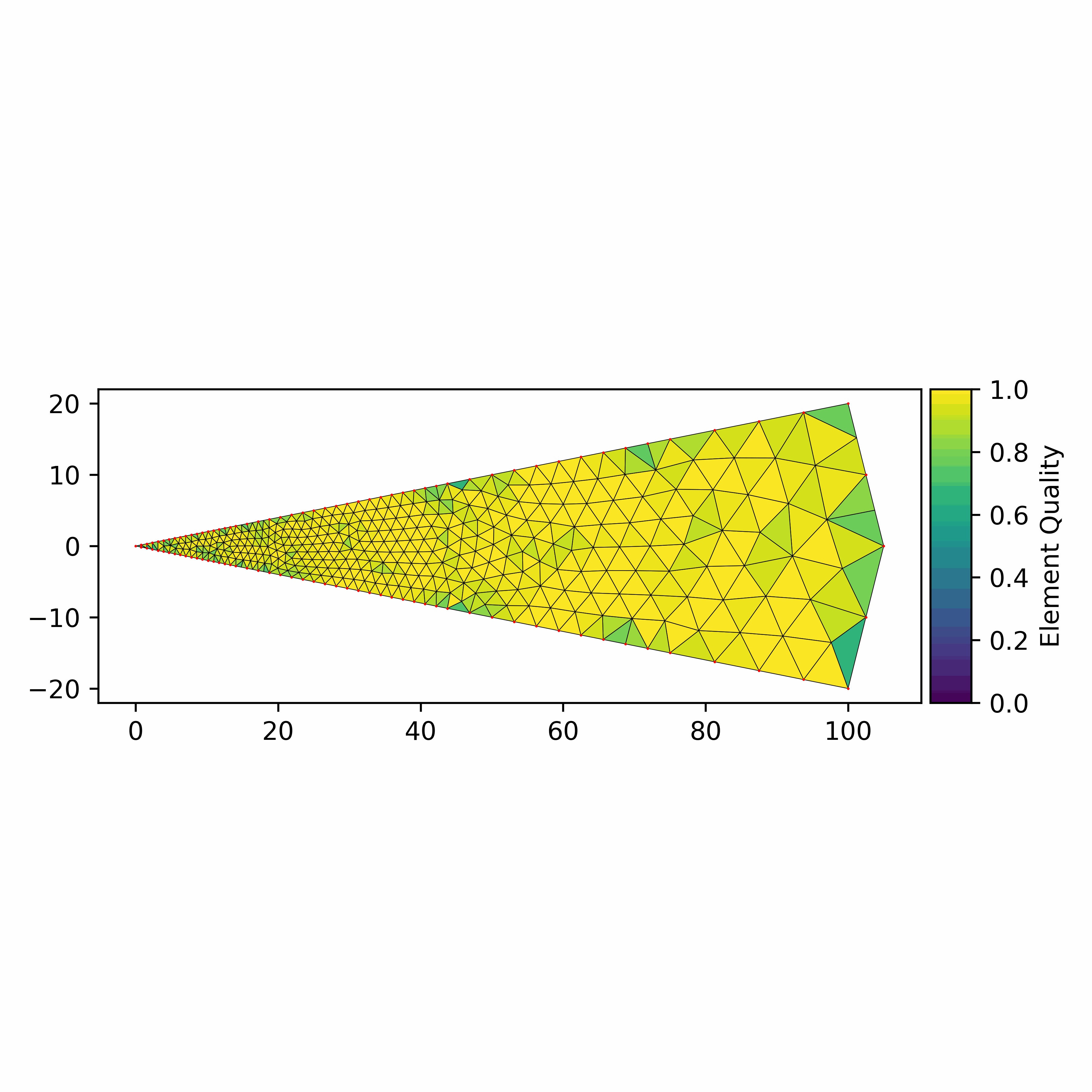}}}
    \hspace{0mm}
    \subfloat{{\includegraphics[trim={50 1100 40 1200},clip,width=.5\linewidth,height=6.5cm,keepaspectratio]{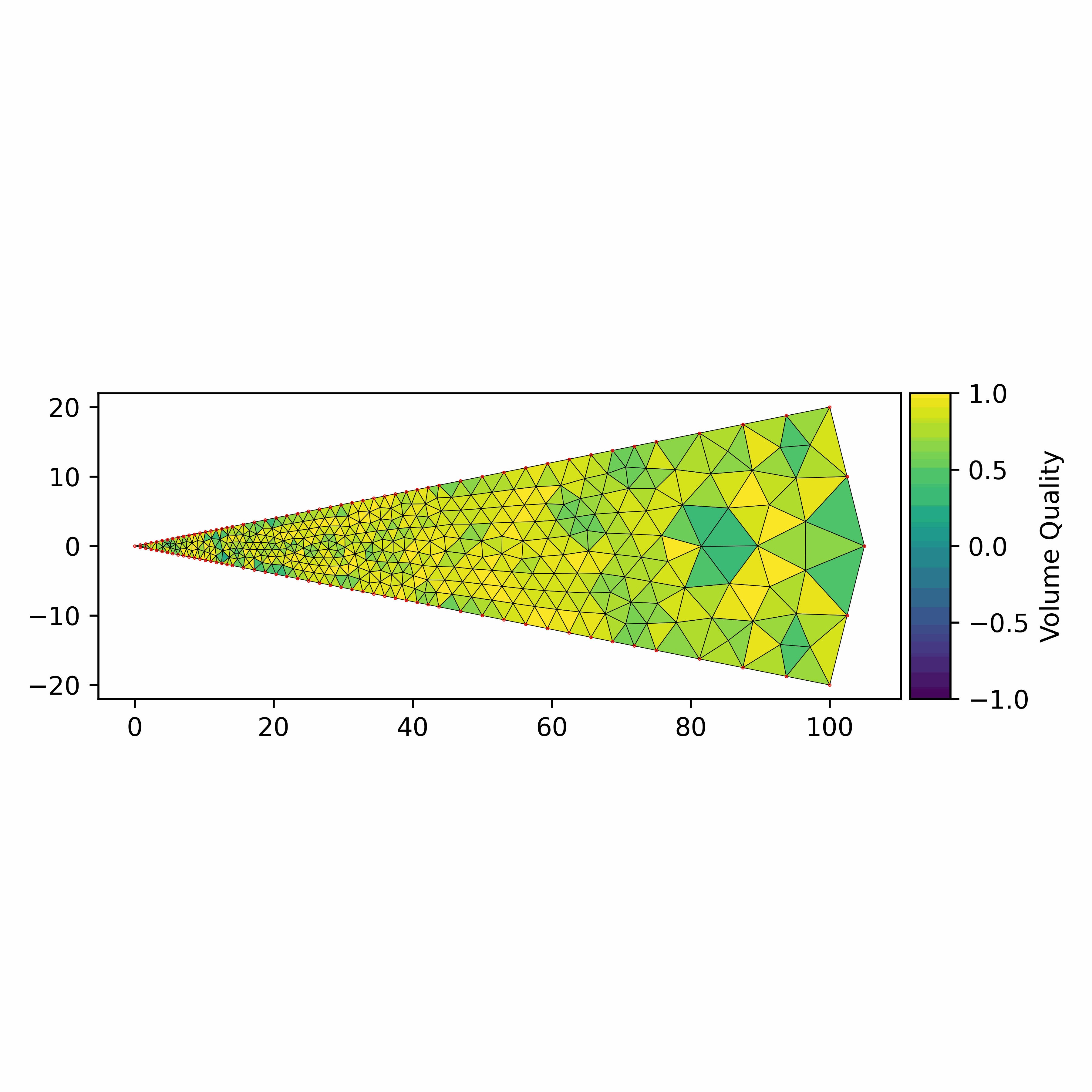}}}
    \subfloat{{\includegraphics[trim={50 1100 40 1200},clip,width=.5\linewidth,height=6.5cm,keepaspectratio]{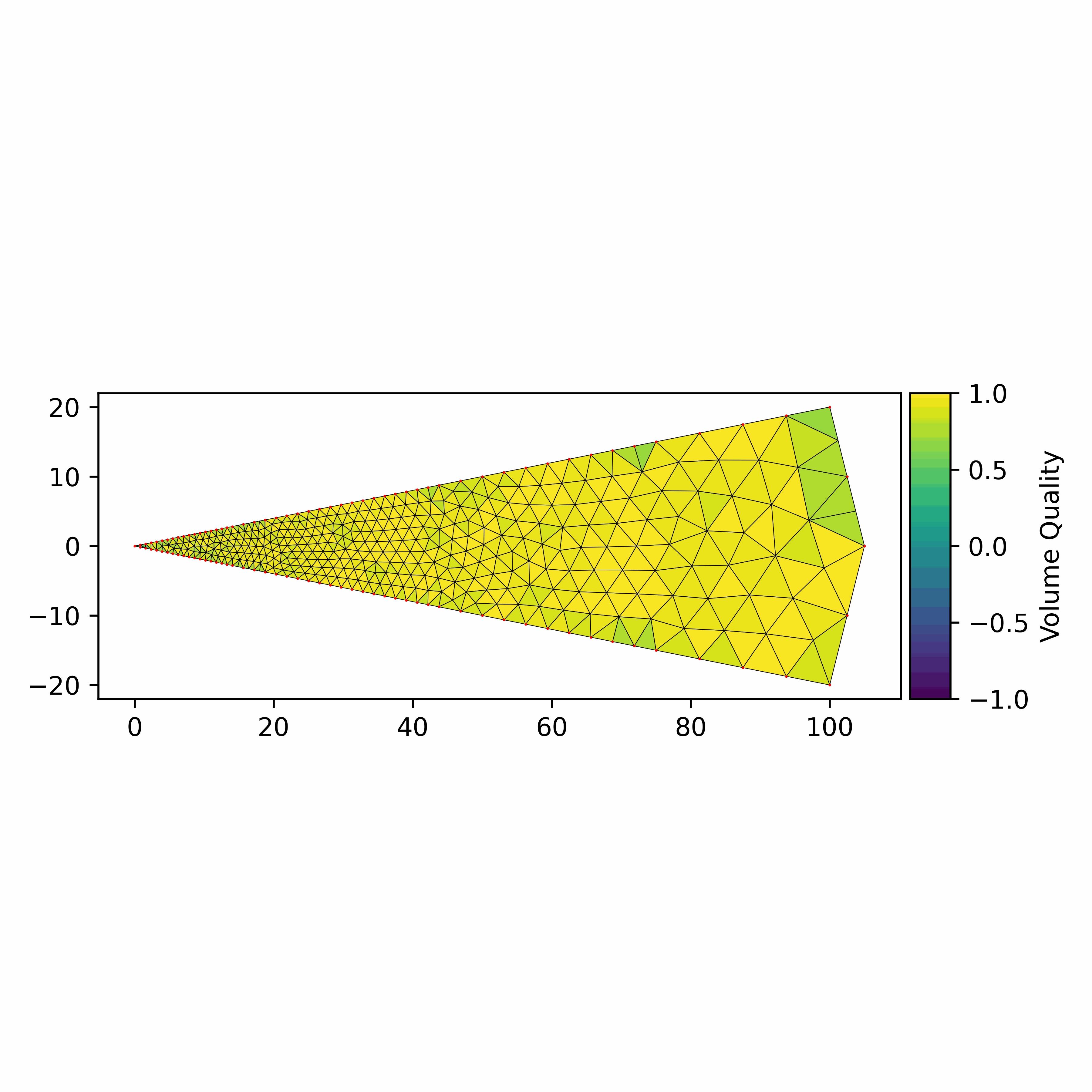}}}
    \hspace{0mm}
    \subfloat{{\includegraphics[trim={50 300 40 350},clip,width=.5\linewidth,height=6.5cm,keepaspectratio]{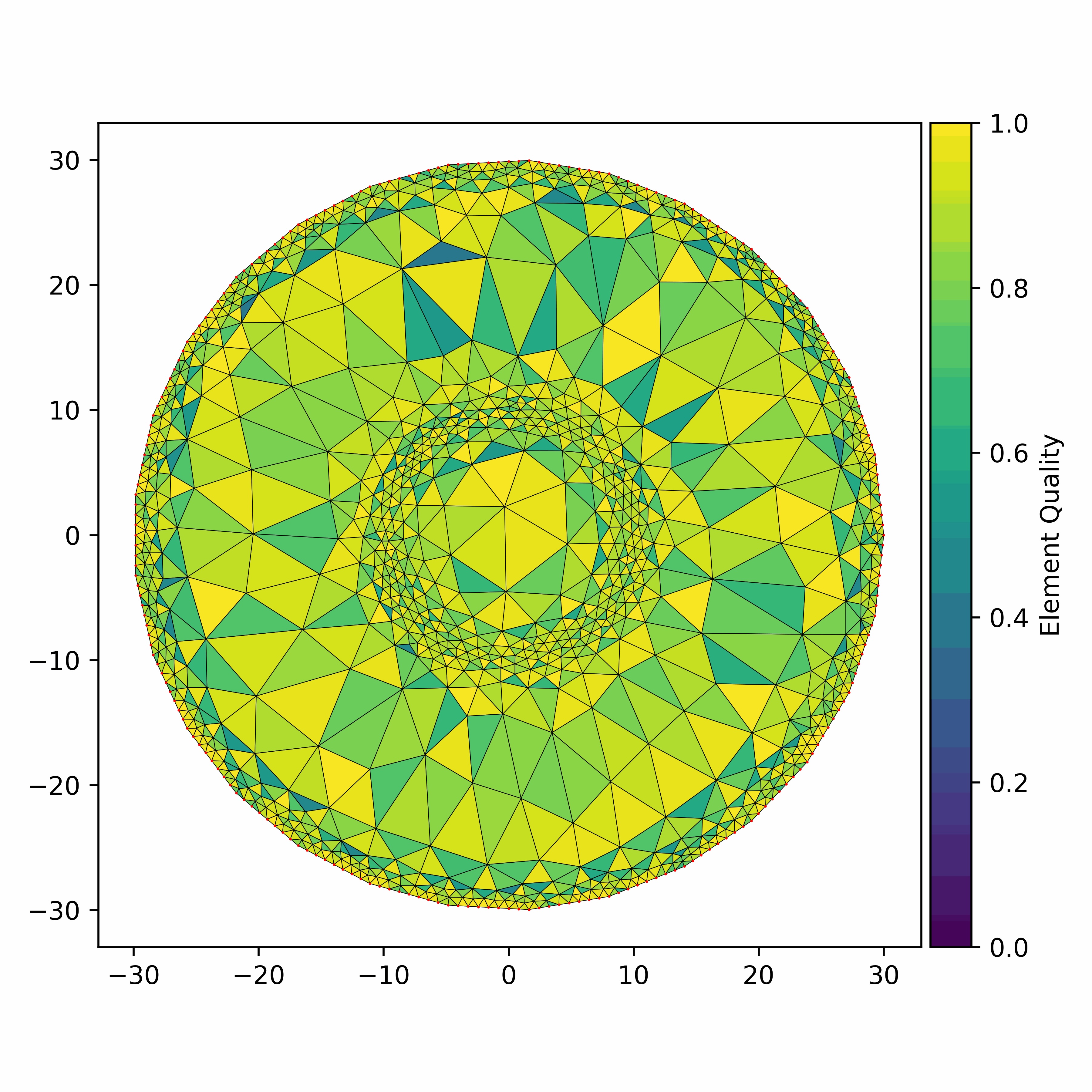}}}
    \subfloat{{\includegraphics[trim={50 300 40 350},clip,width=.5\linewidth,height=6.5cm,keepaspectratio]{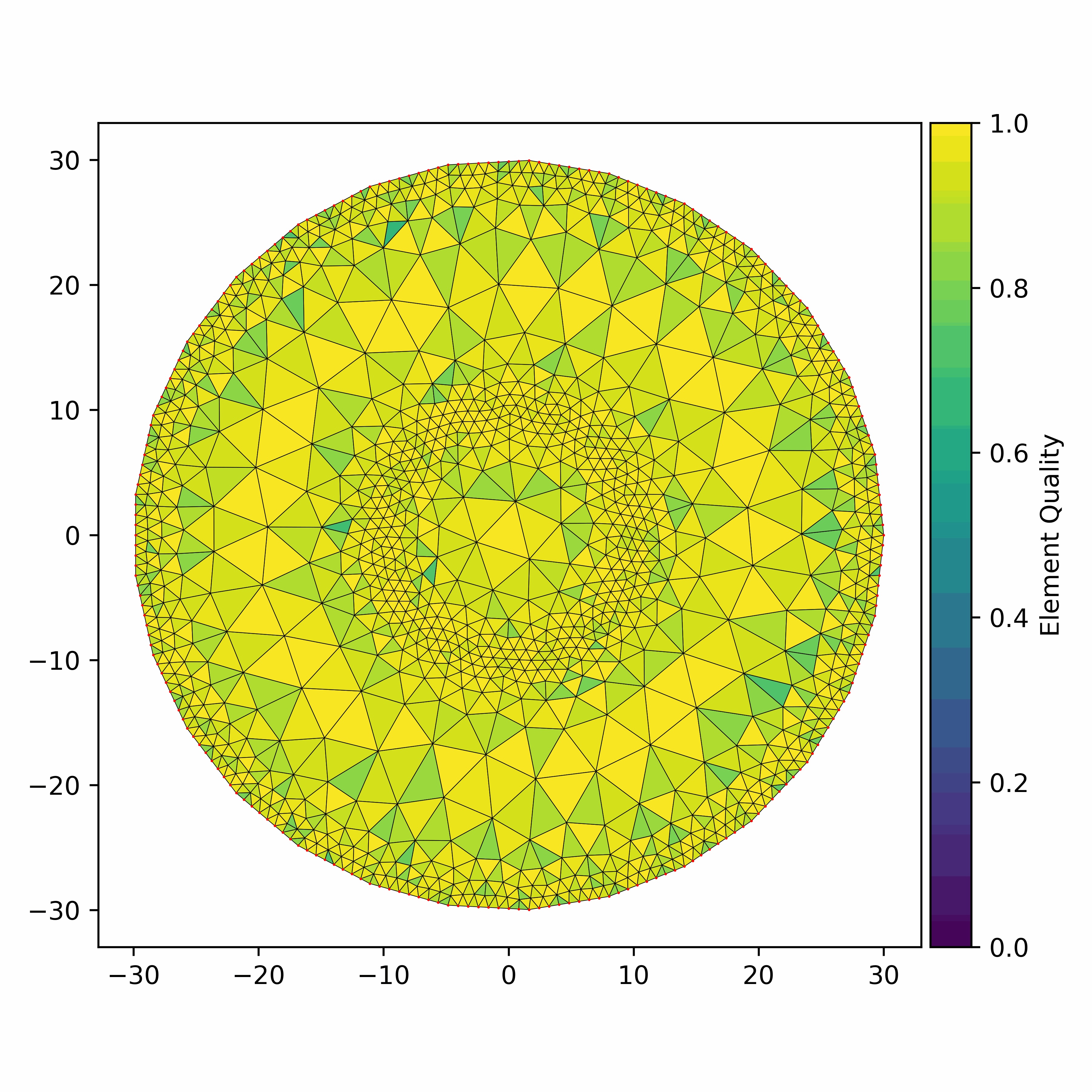}}}
    \hspace{0mm}
    \subfloat{{\includegraphics[trim={50 300 40 350},clip,width=.5\linewidth,height=6.5cm,keepaspectratio]{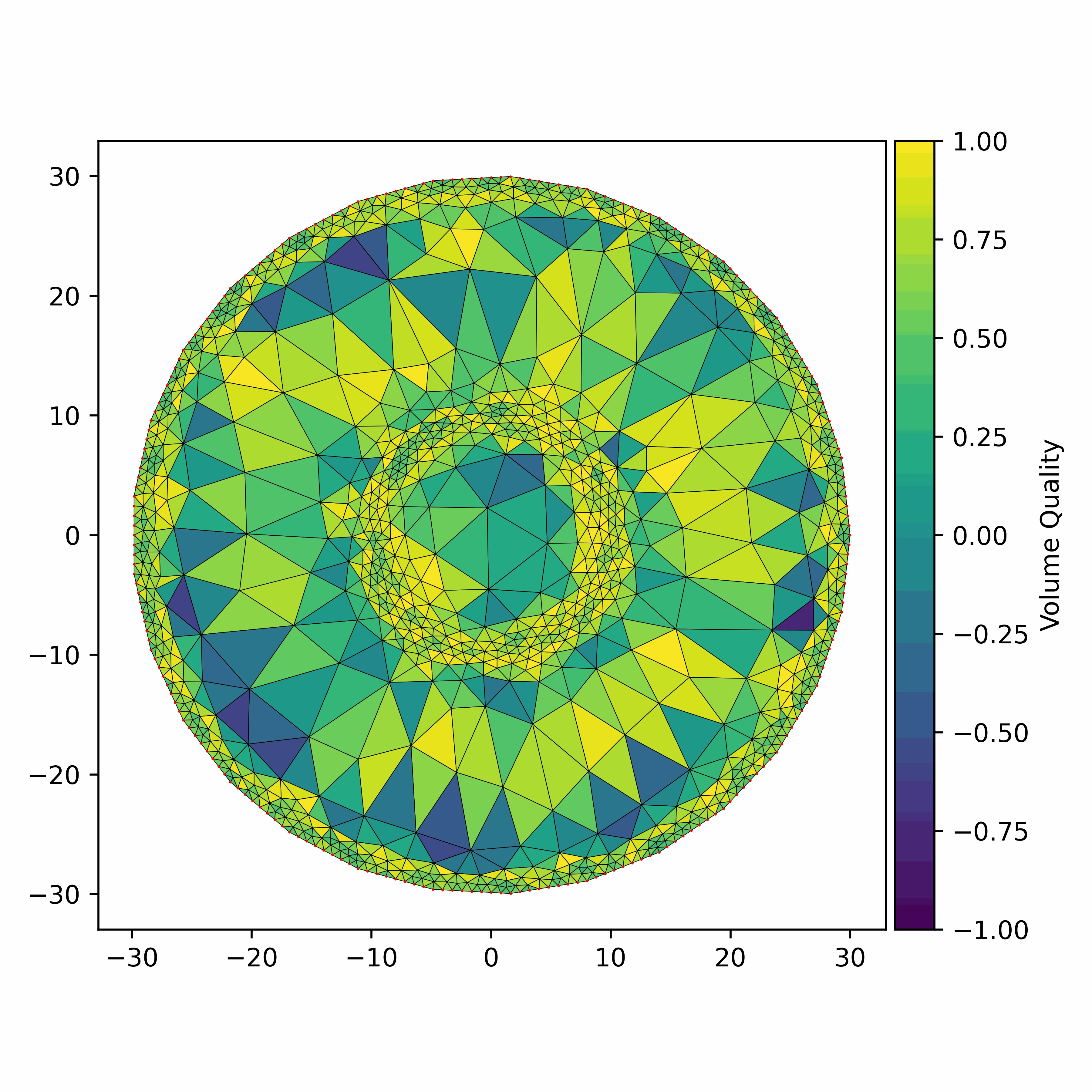}}}
    \subfloat{{\includegraphics[trim={50 300 40 350},clip,width=.5\linewidth,height=6.5cm,keepaspectratio]{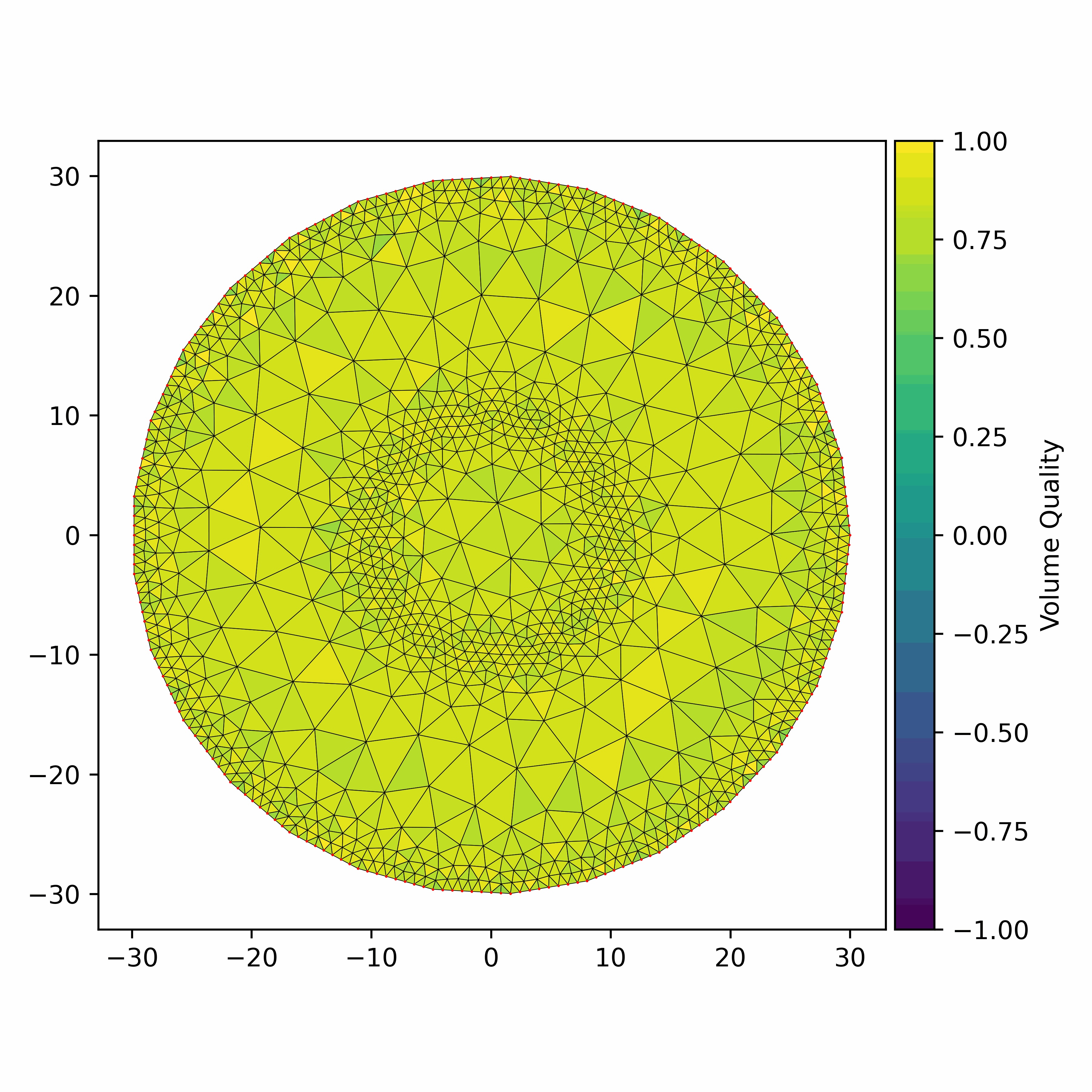}}}
    \caption{Variable resolution meshes. Left hand meshes are produced by trained mesh generator, right hand meshes are produced by DistMesh.}
    \label{fig:h-size-meshes}
\end{figure}

Figure \ref{fig:h-size-meshes} shows \deleted[id=0]{two} \added[id=R1]{three} examples:
\begin{enumerate}
    \item Chip shape with $h(x,y) = 1 + \frac{x^2 + y^2}{40^2}$
    \item Circle shape with $h(x,y) = 5 - 4 \sin^2 \left(\frac{\pi \sqrt{x^2+y^2}}{20} \right)$
    \item \added[id=R1]{NACA0012 Airfoil. See below for size function.}
\end{enumerate}
When the size function changes rapidly, as in the second example, the mesh will take longer to converge to a final state because as we add more nodes we obtain more detailed information about the size function. For these runs we use trajectories of length 30. \added[id=R1]{For the airfoil test, we use the geometry and size function specified for the ``NACA0012 airfoil" test at \cite{distmesh_site}. We modify two parameters for this example, setting $h_{lead} = .04$ and $h_{tail}=.08$. To generate the boundary vertices for the trained mesh generator, we use the boundary vertices outputted by running DistMesh on this example.}

\begin{figure}
    \centering
    \subfloat{{\includegraphics[trim={10 30 10 30},clip,width=.5\linewidth,height=6.5cm,keepaspectratio]{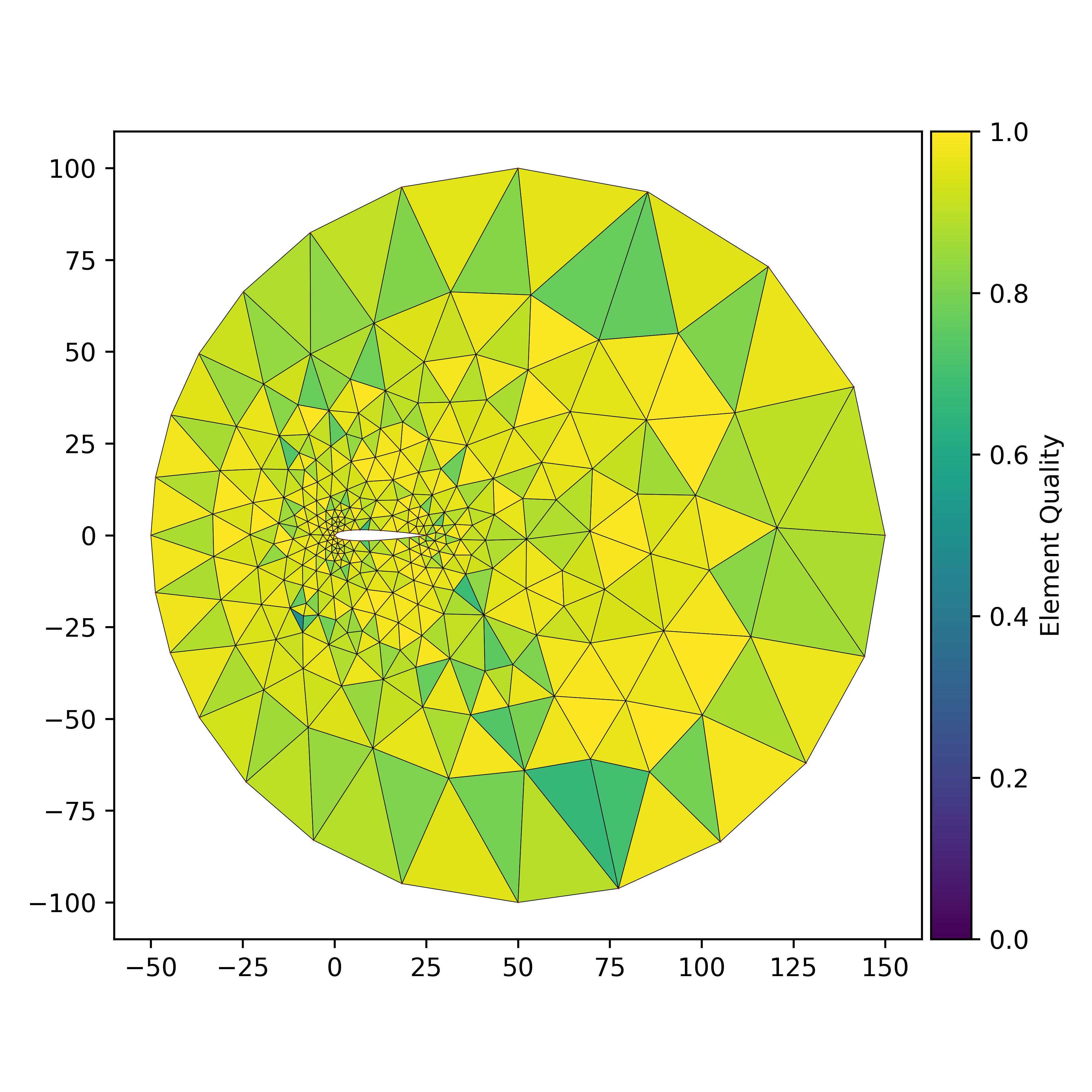}}}
    \subfloat{{\includegraphics[trim={10 30 10 30},clip,width=.5\linewidth,height=6.5cm,keepaspectratio]{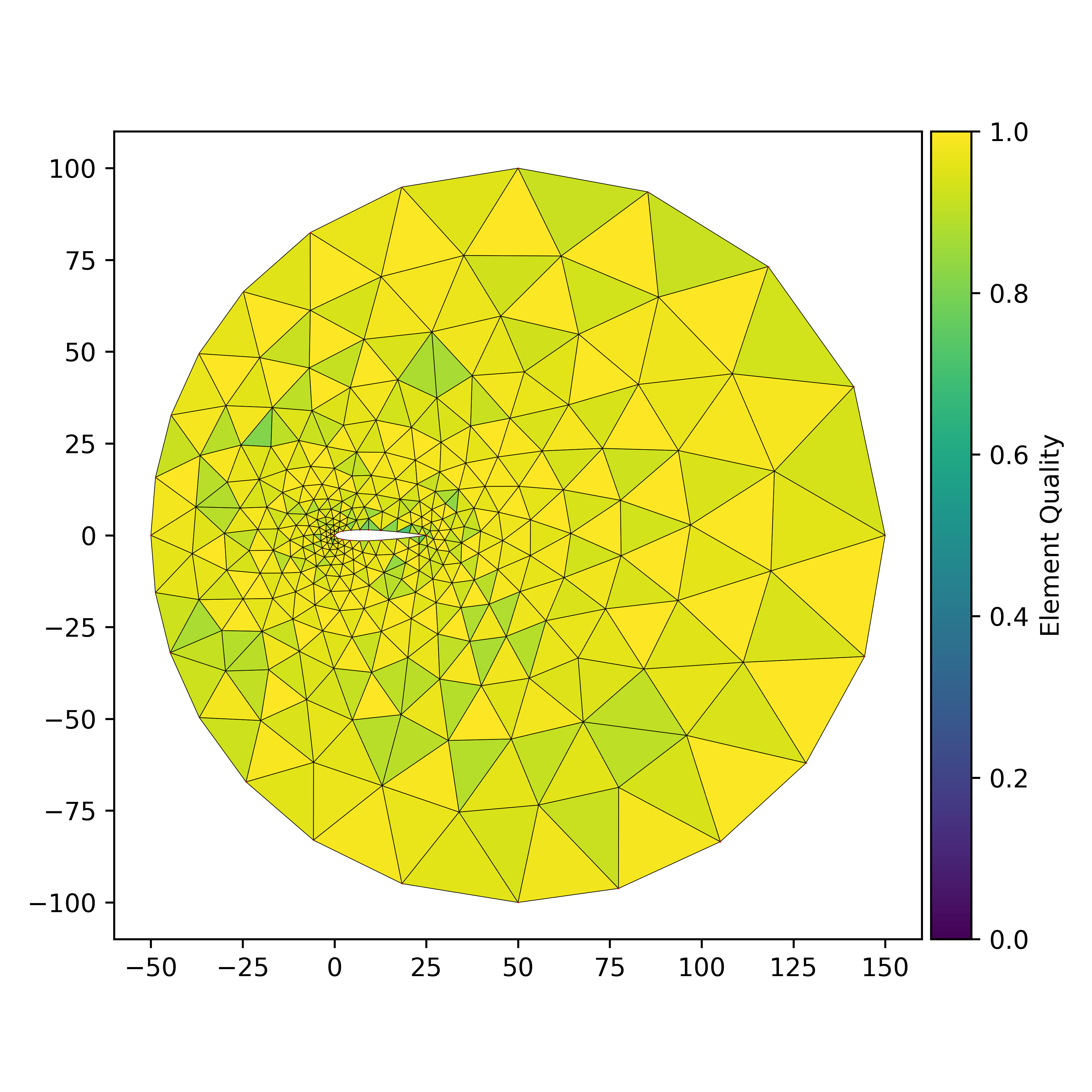}}}
    \hspace{0mm}
    \subfloat{{\includegraphics[trim={10 30 10 30},clip,width=.5\linewidth,height=6.5cm,keepaspectratio]{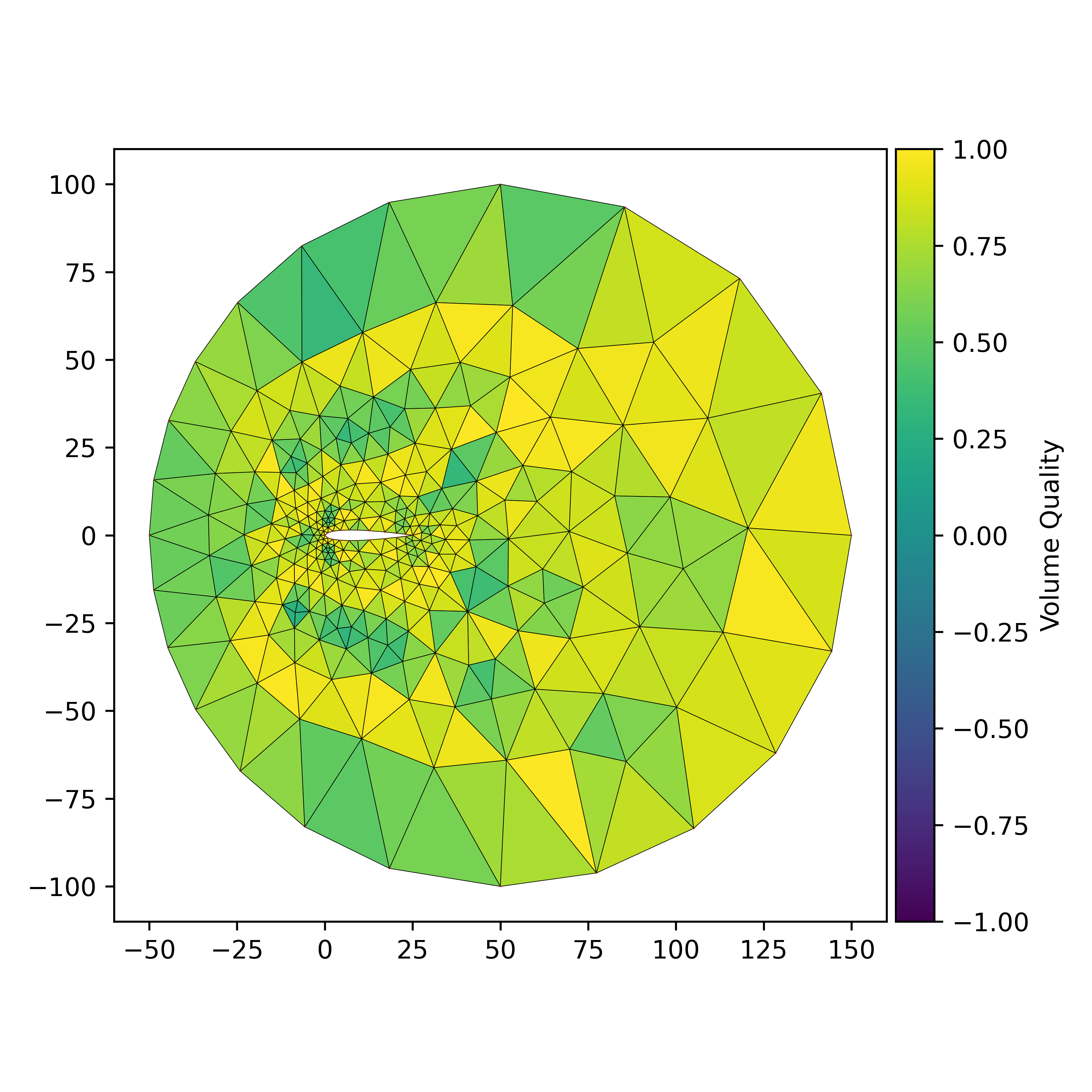}}}
    \subfloat{{\includegraphics[trim={10 30 10 30},clip,clip,width=.5\linewidth,height=6.5cm,keepaspectratio]{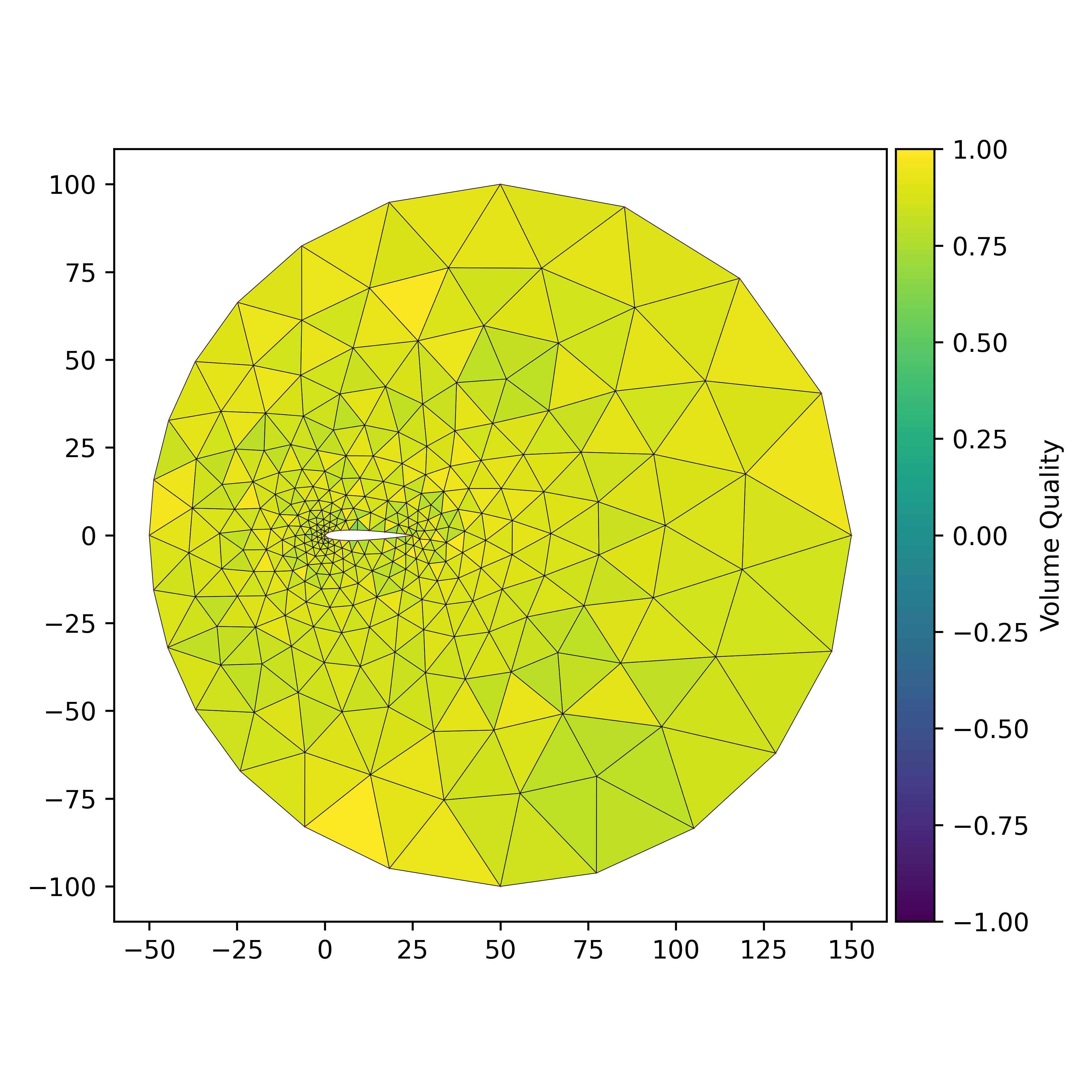}}}
    \hspace{0mm}
    \subfloat{{\includegraphics[clip,width=.475\linewidth,height=5.5cm,keepaspectratio]{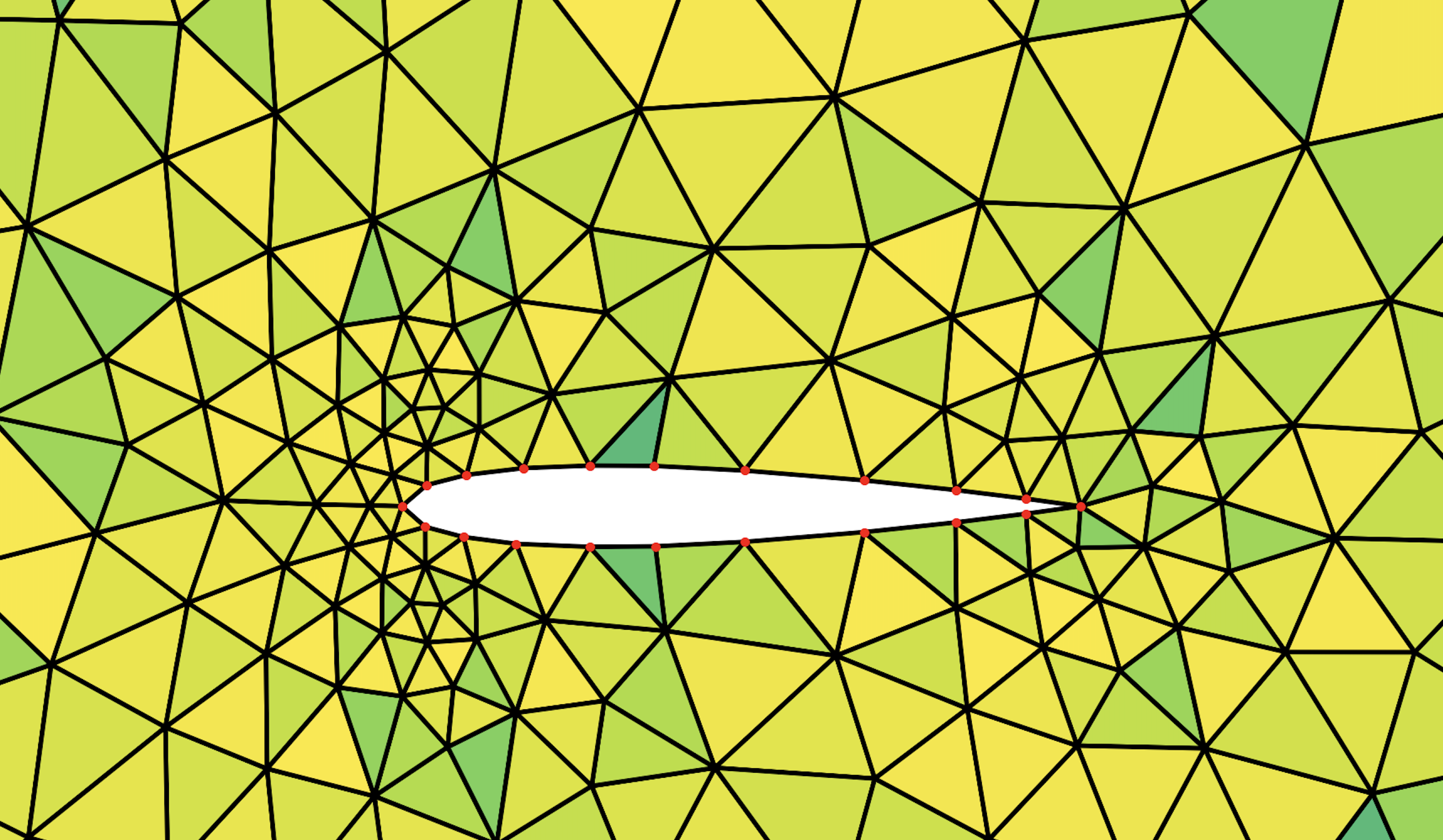}}}
    \hspace{0.025\textwidth}
\subfloat{{\includegraphics[clip,width=.475\linewidth,height=5.5cm,keepaspectratio]{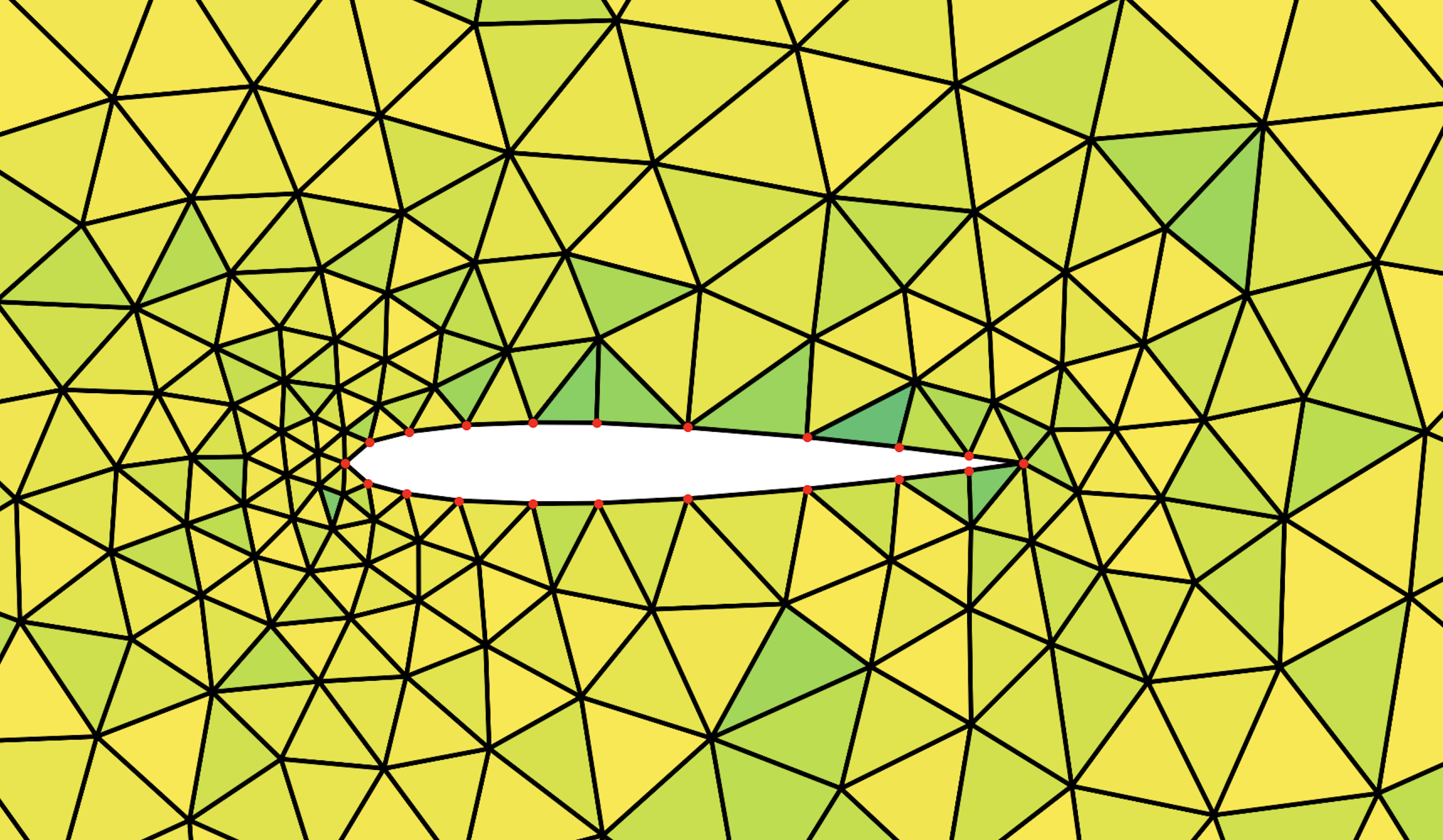}}}

    \caption{\added[id=R1]{NACA Airfoil Test. Left hand meshes are produced by trained mesh generator, right hand meshes are produced by DistMesh. Bottom panels show element quality around airfoil.}}
    \label{fig:naca-test}
\end{figure}

Mesh quality metrics for these tests are shown in Table \ref{tab:hsize_tab}, and meshes are displayed in Figure \ref{fig:h-size-meshes}. These metrics are adjusted for the size function, meaning we use the value of the size function at the midpoint or centroid of an edge or volume, respectively, to calculate target edge lengths and volumes. For the first test, the trained mesh generator achieves similar results to DistMesh in terms of angle and element qualities, but does not produce edges and volumes that are of as high quality. For the second case, DistMesh performs better across all metrics. DistMesh produces elements that are uniformly of high quality, as seen by the high minimum element quality and low standard deviation, and element sizes vary smoothly. The trained mesh generator produces some skinny elements and some elements that do not have the desired volume, and the transition between elements of different sizes is sharper. \added[id=R1]{The NACA airfoil test evaluates our mesh generator's performance on a practical engineering geometry. DistMesh again produces performs better across all categories, in particular the volume and edge quality metrics. As shown in Figure \ref{fig:naca-test}, the trained mesh generator is able to create small elements near the airfoil boundary, but does not make as smooth a transition to coarser elements as we move away from the airfoil boundary as DistMesh does. We can see that the trained mesh generator produces elements that are mostly of high quality (above $.9$ on average), but there are 5-10 misshapen elements scattered throughout the mesh.} While our trained mesh generator achieves comparable results on simple size functions, improving performance on more complex size functions will be the subject of future research.

\begin{table}[h!]
\centering
\resizebox{\textwidth}{!}{\begin{tabular}{|c|c|c|c|c|c|c|c|c|c|c|c|c|c|}
\hline
    &$q_a$ (Mean) & $q_a$ (Min) & $q_a$ (SD) & $q_e$ (Mean) & $q_e$ (Min) & $q_e$ (SD) & $q_r$ (Mean) & $q_r$ (Min) & $q_r$ (SD) & $q_v$ (Mean) & $q_v$ (Min) & $q_v$ (SD) \\ \hline
    Chip (RL) & 0.865  & \textbf{0.280}  & 0.108  & 0.881  & 0.349  & 0.089  & 0.953  & 0.520  & 0.056  & 0.809 &0.239  & 0.143   \\ \hline
    Chip (DistMesh) & \textbf{0.866}  & 0.000  & 0.114  & \textbf{0.922}  & \textbf{0.654}  & 0.067  & 0.953  & 0.520  & 0.058  & \textbf{0.924} & \textbf{0.282}  & 0.076   \\ \hline
    Circle (RL) & 0.753  & -0.091  & 0.180  & 0.775  & -0.050  & 0.171  & 0.866  & 0.395  & 0.112  & 0.637 &-0.803  & 0.251   \\ \hline
    Circle (DistMesh) & \textbf{0.843}  & \textbf{0.292}  & 0.117  & \textbf{0.927}  & \textbf{0.712}  & 0.051  & \textbf{0.941}  & \textbf{0.687}  & 0.053  & \textbf{0.897} &\textbf{0.688}  & 0.059  \\ \hline
    \added[id=R1]{Airfoil (RL)} & \added[id=R1]{0.828}  & \added[id=R1]{0.034}  & \added[id=R1]{0.127}  & \added[id=R1]{0.842}  & \added[id=R1]{0.416}  & \added[id=R1]{0.118}  & \added[id=R1]{0.928} & \added[id=R1]{0.490}  & \added[id=R1]{0.065} & \added[id=R1]{0.752} &\added[id=R1]{0.274} & \added[id=R1]{0.176}  \\ \hline 
    \added[id=R1]{Airfoil (DistMesh)} & \added[id=R1]{\textbf{0.866} }  & \added[id=R1]{\textbf{0.361}}  & \added[id=R1]{0.101}   & \added[id=R1]{\textbf{0.920}} & \added[id=R1]{\textbf{0.715}}  & \added[id=R1]{0.054} & \added[id=R1]{\textbf{0.956}}  & \added[id=R1]{\textbf{0.710}}  & \added[id=R1]{0.040} & \added[id=R1]{\textbf{0.868}} &\added[id=R1]{\textbf{0.634}} & \added[id=R1]{0.048}  \\\hline
\end{tabular}}
\caption{Mesh quality metrics for size function test.}
\label{tab:hsize_tab}
\end{table}

\subsection{Mesh Improvement Capability}
In this section we demonstrate that we can train a mesh generator to improve an existing mesh. During training, the initial condition for each trajectory is a mesh with randomly distributed vertices. To be specific, we initialize a uniform grid of equilateral triangles with edge length $2^m$, where $m$ is sampled uniformly from $\{-2, \ldots ,2\}$, and then each vertex position is perturbed by a random sample from a zero mean normal distribution with variance equal to the edge length. We use this initialization scheme during training so that our mesh generator learns a generalizable strategy for improving meshes. We use the same architecture and training curriculum as prior experiments, and our reward function is the same as the baseline model. For this experiment we found better results using a deterministic policy.

In our first experiment we initialize a uniform grid of equilateral triangles, and remove vertices that lie outside the polygon. From Table \ref{tab:init_comp} we see that with such an initialization we obtain comparable results to DistMesh, which internally uses this same uniform initialization strategy for a constant size function. Our mesh generator produces average and minimum angles that are slightly higher quality, and edges and volumes that are slightly lower quality than DistMesh.

Next we examine the trained mesh generator's ability to improve meshes produced by Triangle and DistMesh. For DistMesh, angle and element qualities are slightly improved, and edge and volume qualities are slightly lessened. For Triangle, average angle and element qualities are improved by about ten percent, and the minimum angle quality is, on average, about doubled. 

\begin{figure}
    \centering
    \subfloat{{\includegraphics[trim={110 500 100 600},clip,width=0.5\linewidth]{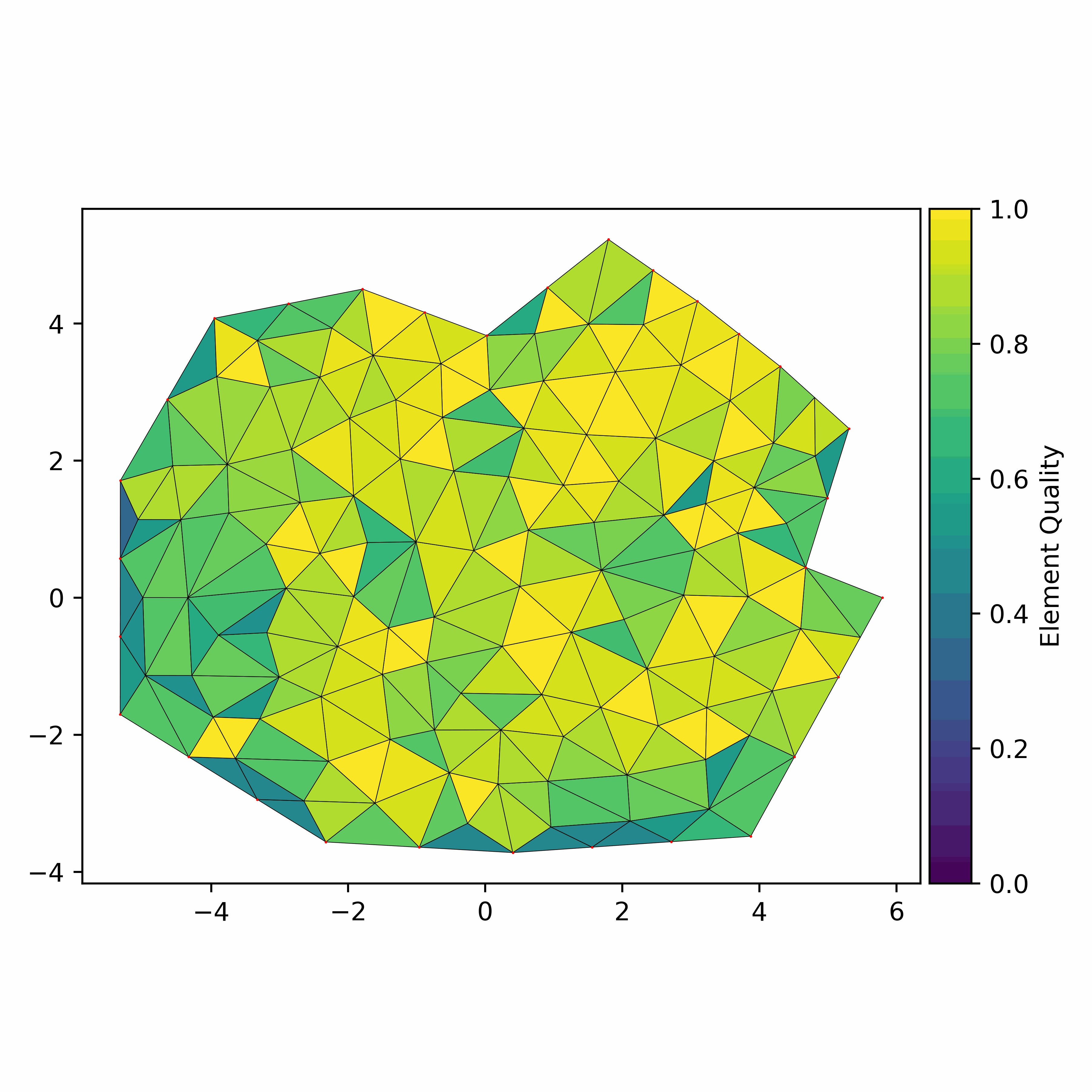}}}
    \subfloat{\includegraphics[trim={110 500 100 600},clip,width=0.5\linewidth]{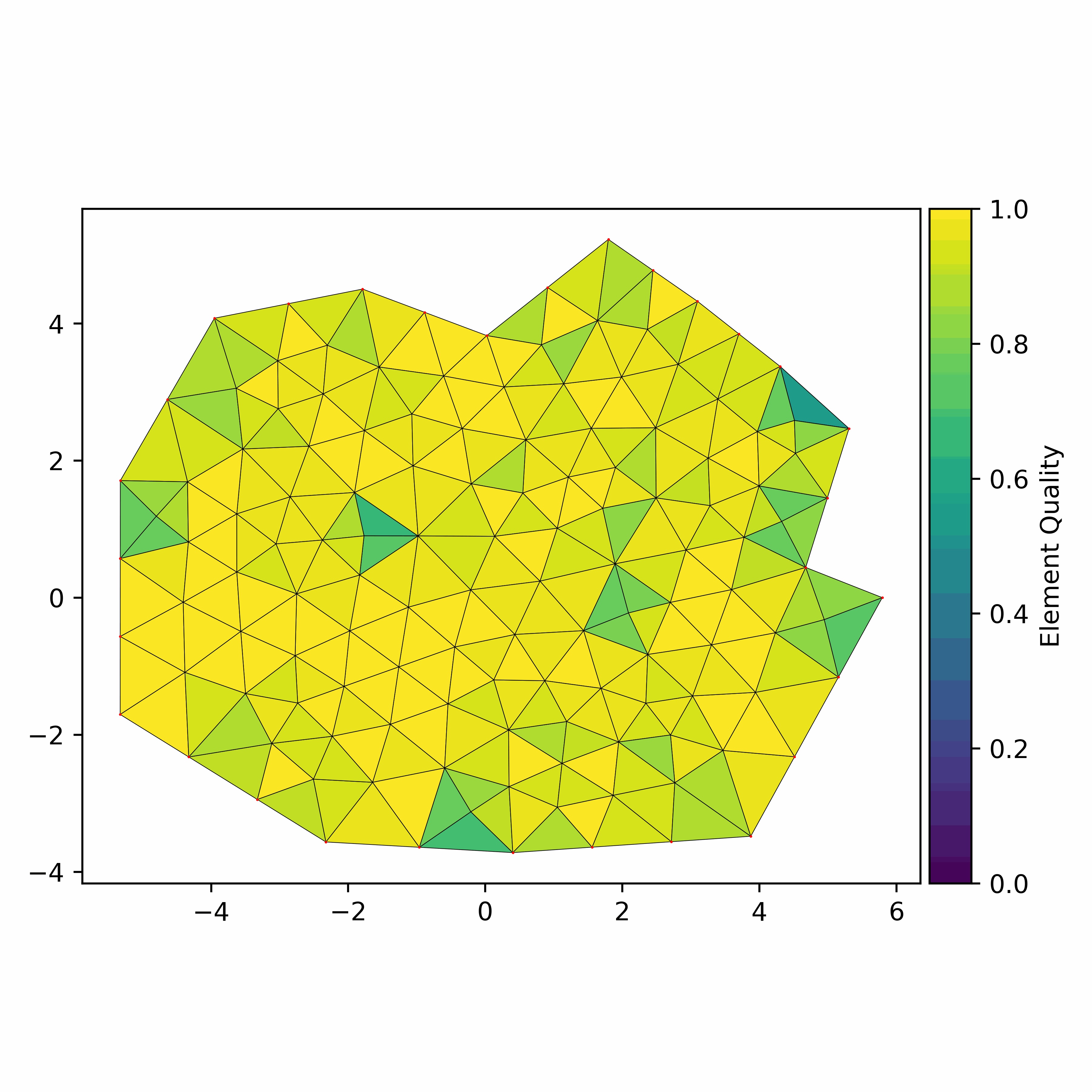}}
    \hspace{0mm}
    \subfloat{{\includegraphics[trim={110 500 100 600},clip,width=0.5\linewidth]{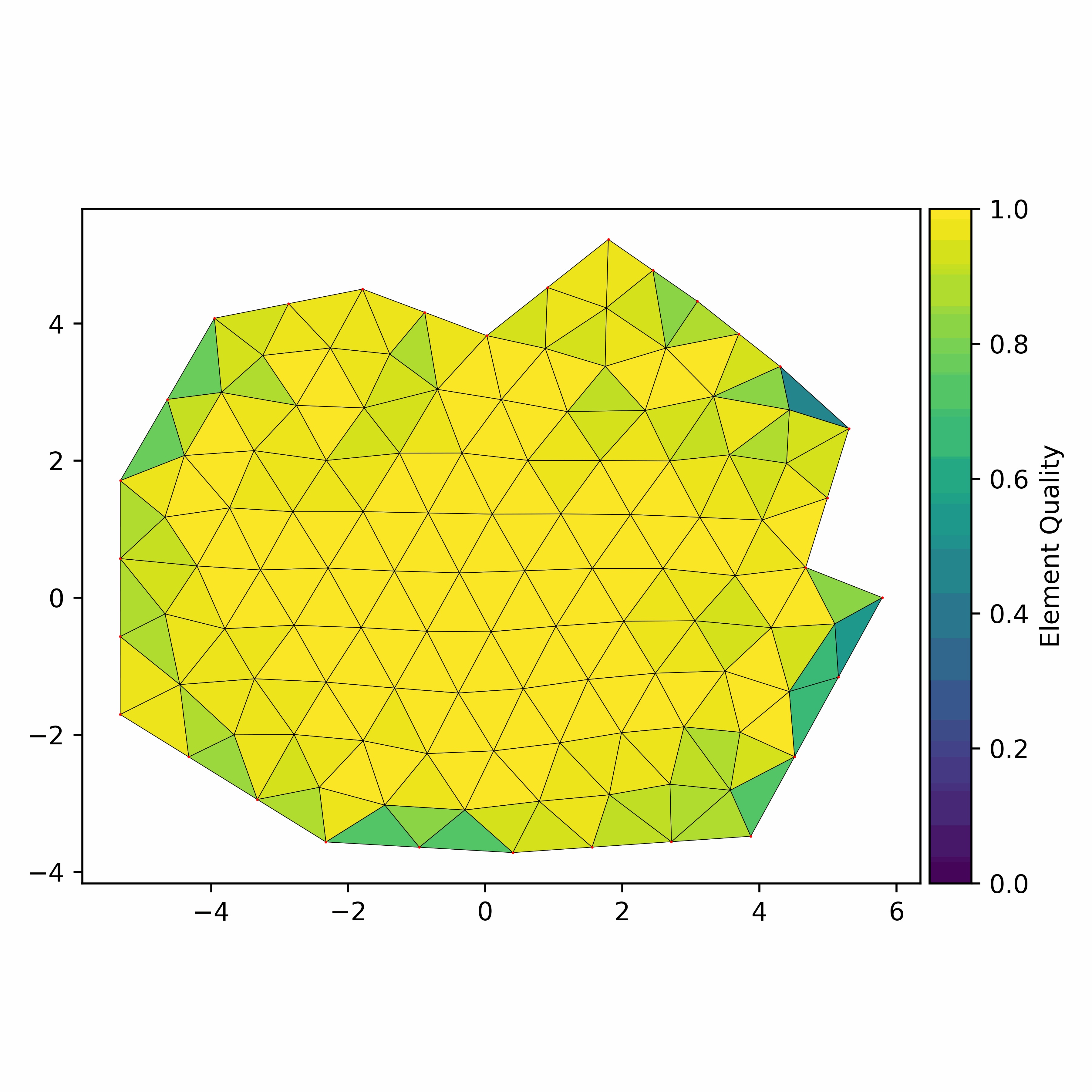}}}
    \subfloat{\includegraphics[trim={110 500 100 600},clip,width=0.5\linewidth]{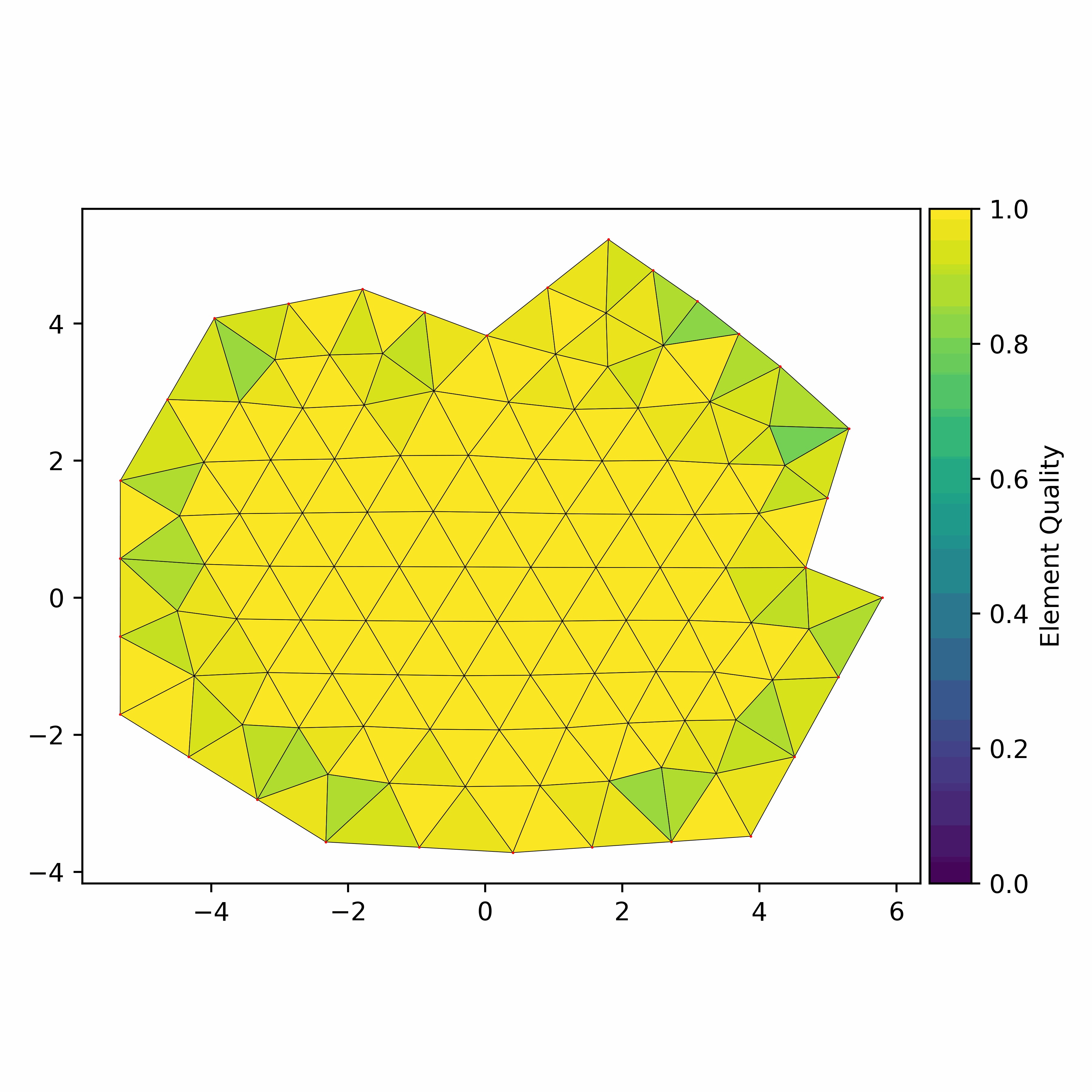}}
    \hspace{0mm}
    \subfloat{{\includegraphics[trim={110 500 100 600},clip,width=0.5\linewidth]{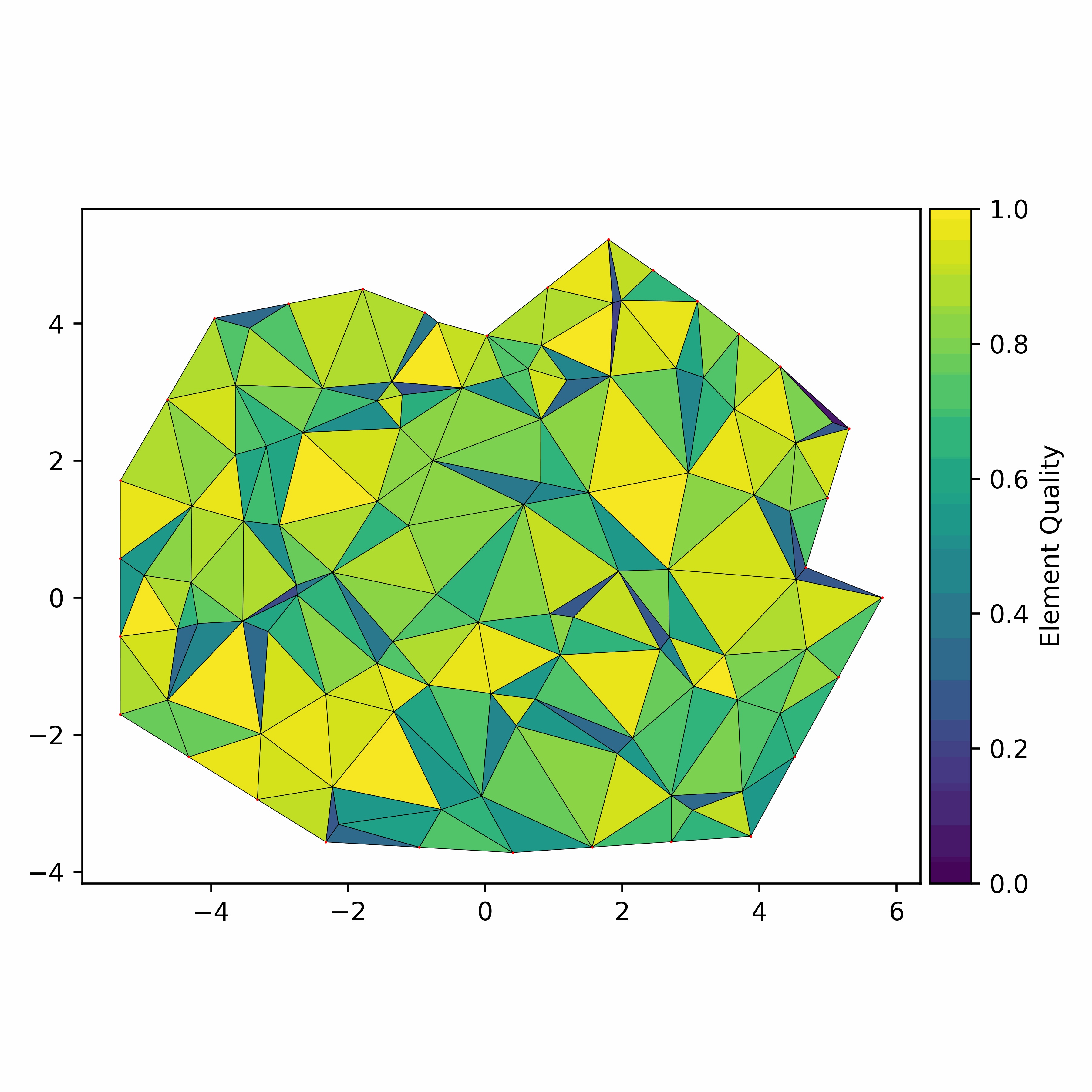}}}
    \subfloat{\includegraphics[trim={110 500 100 600},clip,width=0.5\linewidth]{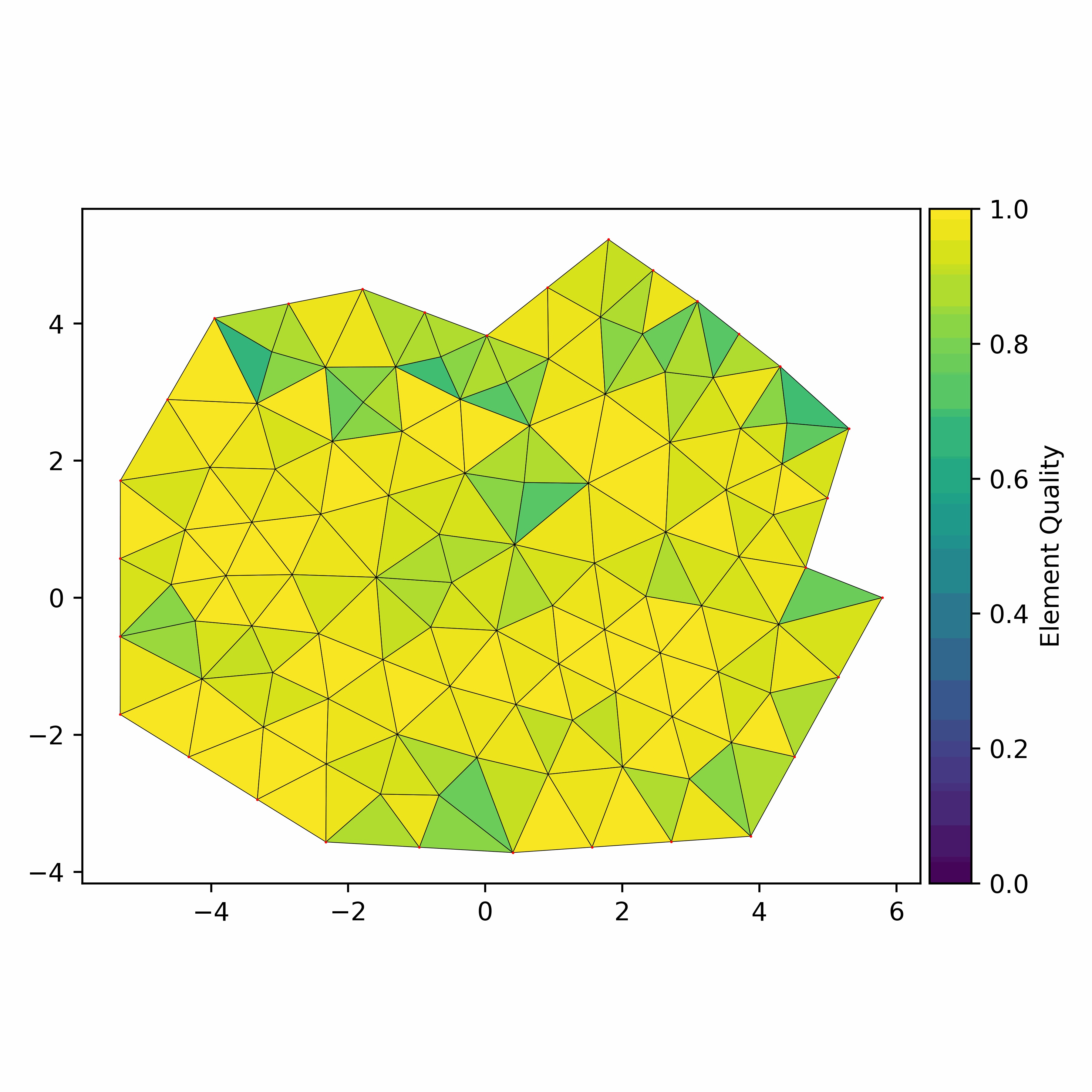}}
    \caption{Mesh improvement examples, where left side is initial mesh and right is final mesh.  Top Row: Triangle initialization. Middle: DistMesh initialization Bottom: Randomly perturbed uniform equilateral grid initialization.}
    \label{fig:mesh_imp}
\end{figure}

In Figure \ref{fig:mesh_imp} we show initial and final meshes after the improvement process. For both DistMesh and Triangle, low quality elements near the boundary are improved by moving vertices inward and re-shaping those elements. We additionally show an example of improving a randomly perturbed uniform equilateral mesh\added[id=R1]{, which initially contains highly distorted elements}. We see that sliver elements as well as very small elements are removed by deleting and moving vertices. All elements in the improved mesh, for this particular example, have element quality of greater than $.65$. We note that our mesh generator is able to improve meshes by simply moving, adding and deleting nodes, while some existing mesh improvement algorithms \cite{https://doi.org/10.1002/(SICI)1097-0207(19971115)40:21<3979::AID-NME251>3.0.CO;2-9}, \cite{Klinger} additionally use techniques like edge-swapping.

\begin{table}[h!]
\centering
\resizebox{\textwidth}{!}{\begin{tabular}{|c|c|c|c|c|c|c|c|c|c|c|c|c|}
\hline
 &$q_a$ (Mean) & $q_a$ (Min) & $q_a$ (SD) & $q_e$ (Mean) & $q_e$ (Min) & $q_e$ (SD) & $q_r$ (Mean) & $q_r$ (Min) & $q_r$ (SD) & $q_v$ (Mean) & $q_v$ (Min) & $q_v$ (SD) \\ \hline
RL (No init.) & 0.860  & \textbf{0.198}  & 0.118  & 0.862  & 0.327  & 0.114  & 0.948  & 0.619  & 0.060  & 0.779 &-0.127  & 0.182   \\ \hline
RL (Unif. Init.) & 0.924 & 0.194 & 0.110 & 0.923 & 0.503 & 0.076 & 0.973 & \textbf{0.586} & 0.056 & 0.870 & 0.379 & 0.109 \\ \hline
RL (Triangle Init.) & 0.853 & 0.110 & 0.119 & 0.829 & 0.457 & 0.109 & 0.945 & 0.543 & 0.060 & 0.706 & 0.290 & 0.158 \\ \hline
RL (DistMesh Init.) & \textbf{0.926} & 0.129 & 0.104 & 0.925 & 0.554 & 0.068 & \textbf{0.976} & 0.565 & 0.053 & 0.873 & 0.421 & 0.099 \\ \hline
Triangle & 0.757 & -0.262 & 0.191 & 0.815 & 0.538 & 0.107 & 0.868 & 0.274 & 0.125 & 0.723 & 0.359 & 0.156\\ \hline
DistMesh & 0.903 & 0.004 & 0.112 & \textbf{0.938} & \textbf{0.609} & 0.063 & 0.967 & 0.495 & 0.058 & \textbf{0.926} & \textbf{0.565} & 0.061 \\ \hline
\end{tabular}}
\caption{Mesh quality metrics for mesh improvement test.}
\label{tab:init_comp}
\end{table}

\section{Conclusion}
In this paper we introduced a mesh generator that can be trained to maximize a given mesh quality metric using reinforcement learning. The input to the mesh generator is a list of boundary vertices that define a polygonal domain, and the output is a triangular mesh. The policy function is a convolutional graph neural network which respects rotational and translational symmetry. The policy function can move, add and delete vertices, and the triangulation is determined by the Delaunay algorithm. We train the mesh generator using the PPO algorithm, a policy gradient method that updates the parameters of the stochastic policy to maximize the reward from sampled trajectories. 

This work is a promising proof-of-concept for building a mesh generator using reinforcement learning. We have demonstrated that the mesh generator can be trained on the task of making small uniform meshes of random polygonal domains, and generalize to produce meshes of arbitrary sized and shaped domains, including variable resolution meshes, as well as improve existing meshes. The quality of the meshes is comparable to widely-used Delaunay mesh generators \added[id=R1]{for the task of producing uniform meshes, and it underperforms Distmesh for variable resolution meshes. Improving this capability is a goal for future research.}

There are several \deleted[id=0]{future} \added[id=0]{other potential} research directions. The policy function is agnostic to the spatial dimension and the type of element, as it takes an arbitrary coordinate graph as input, so we could try to extend this framework to higher dimensions or to other element types such as hexahedrons. This will certainly introduce new challenges, as domains and elements become more complex in higher dimensions\added[id=R1]{, and the available triangulation methods may not be as reliable as the two dimensional Delaunay algorithm}. We can experiment with different graph neural network architectures, as well as other reinforcement learning algorithms\deleted[id=0]{ such as Soft-Actor Critic \cite{haarnoja2018softactorcriticoffpolicymaximum}}. We may find it beneficial to use a neural network with a more ``expressive" architecture, as our experiments indicated that we could not increase performance by increasing the size of the network for our chosen architecture. Lastly, we are interested in including the triangulation as part of the policy, rather than relying on an external algorithm; by doing so we could achieve an end-to-end learned mesh generator.









\bibliographystyle{elsarticle-num} 
\bibliography{references}
\end{document}